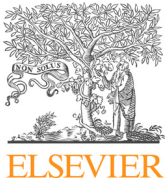
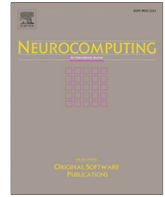

# Hyper-class representation of data

Shichao Zhang [a], Jiaye Li [a,*], Wenzhen Zhang [b], Yongsong Qin [b]

[a] *the School of Computer Science and Engineering, Central South University, Changsha 410083, China*
[b] *Guangxi Key Lab of Multi-Source Information Mining and Security, Guangxi Normal University, Guilin, Guangxi 541004, PR China*

**ABSTRACT**

Data representation is usually a natural form with their attribute values. On this basis, data processing is an attribute-centered calculation. However, there are three limitations in the attribute-centered calculation, saying, inflexible calculation, preference computation, and unsatisfactory output. To attempt the issues, a new data representation, named as hyper-classes representation, is proposed for improving recommendation. First, the cross entropy, KL divergence and JS divergence of features in data are defined. And then, the hyper-classes in data can be discovered with these three parameters. Finally, a kind of recommendation algorithm is used to evaluate the proposed hyper-class representation of data, and shows that the hyper-class representation is able to provide truly useful reference information for recommendation systems and makes recommendations much better than existing algorithms, *i.e.,* this approach is efficient and promising.

© 2022 The Author(s). Published by Elsevier B.V. This is an open access article under the CC BY-NC-ND license (http://creativecommons.org/licenses/by-nc-nd/4.0/).

## 1. Introduction

Data is the core research object in many fields, such as data science, data engineering, data management, database, data mining and big data, aiming to improve data utilization [1–4]. This leads to some key basic topics, mainly including data structure, data organization, data modelling and data representation [5,6]. Among extant data models, relation model should be the most natural data representation form with their attribute values, which makes diverse data utilizations much easy. For example, data enquiry and data mining are two common utilizations of data. Both of them need to compute the distance between two data points. It is relation model that provides convenient information for calculating the distance with Euclidean distance function [7,8].

To better utilize the data, one can weight data attributes for better measuring the intimacy among data, and weight data points for data mining [9,10]. To this end, one may need to introduce a super/hyper variable, called weighting variable. Clearly, drawing hyper variable into distance function is a pretty good way of toning the importance of data attributes (or data points) in real applications. While the hyper variables work well for adjusting the data attributes (or data points), a hyper class (or super class) representation is advocated for data analysis applications, denoted as HC (or SC) representation. It provides a different way of reorganizing attribute values for efficiently utilizing data. This efficiency can easily be showed with tree data structures (see Fig. 1). For example, in a supermarket, data are stored in the form of transaction by transaction with time stamp. These data are often ordered day by day for checking daily profits, as well as month by month for checking monthly profits. To discover trend/fashion patterns in the data, one can organize data season by season, or year by year. In above process, the value of time attribute is changed from infinity to 365, 365 to 12, and 12 to 4. They are illustrated in Fig. 1.

Apparently, HC representation delivers a simple tree data structure in real applications. Another reason of exploring data representation is to provide a way of addressing the unfair operations, such as weighting, to the attributes of data. In weighting process, data attributes are always taken as the minimum operation unit. Once an attribute, $X$, is assigned to a weight $Wi$, no matter what value $X$ is taken, the value wins the weight $Wi$. This is surely unfair in real applications. The reason is that assigning weights to an attribute is based on scoring the correlation between the attribute and decision attribute in training data. Generally, if there are more than 70% samples satisfied a score function, it is referred to a nice score function. In other words, there is a little bit unfairness in score functions. To solve this unfairness, the scoring functions can be constructed as piecewise functions with the HC representation.

From the above, current data processing is an attribute-centered calculation with three limitations, saying, inflexible calculation, preference computation, and unsatisfactory output. Inflexible calculation indicates that attribute-centered calculation is a mechanical process that restricts the performance of some calculations. Consider the height attribute, it is necessary to calculate the difference between the real height values of two people, called as quantitative calculation. However, in some real applications, ones only need to qualitatively know the status of high, medium and low, that is, the height difference between two people in the same height is 0. This is referred to qualitative calculation. Similarly, the age attribute can also be calculated quantitatively and qualitatively according to the need of practical applications. This flexible computing ability is human wisdom, and the existing data calculations can not reflect this flexibility, *i.e.*, it is a inflexible calculation. Preference computation means that the results of data calculation can only meet the needs of normal application. For abnormal scenarios, the results of data calculation cannot fully fit the needs of practical application. In the above, weighting computation is the best example of preference computation. Unsatisfactory output means that the calculation results are mixed, confused and difficult to understand. This limitation can easily be explained by the stock results. For example, there is no law when the conclusion of the Shanghai stock index is obtained with the time-sharing index, whereas it is clear up or down when the conclusion is computed with the annual index. In a word, the three limitations are caused by the inflexibility of attribute-centered calculation.

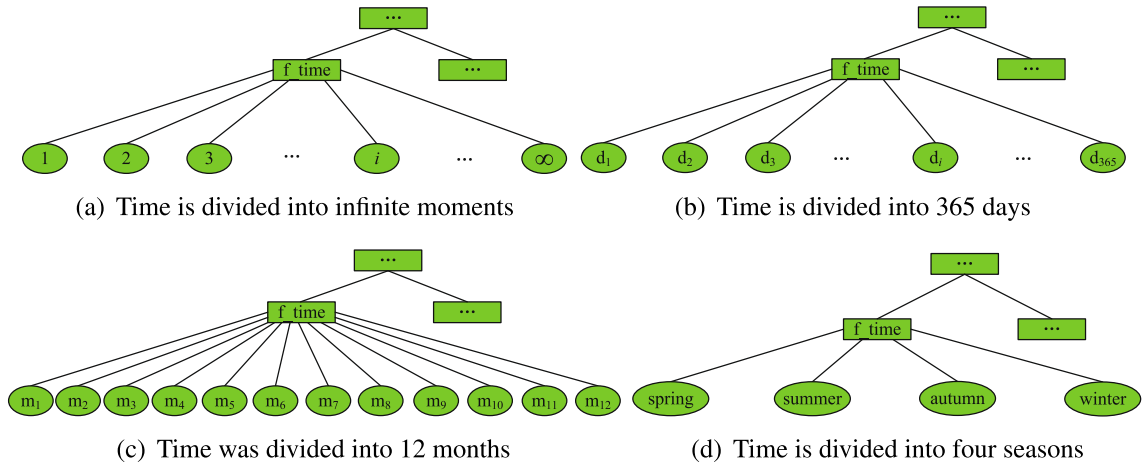

Fig. 1. Different divisions of time.

In this paper, hyper-class representation of data is established by defining Cross Entropy (CE), Kullback–Leibler (KL) divergence and Jensen-Shannon (JS) divergence of features in data. Specifically, a partition of the dataset is generated with each feature, resulting in a total of $d$ partitions (assuming that the data have limited of $d$ features). And then, the remaining $d$-1 features are used to generate $d$ partitions (each feature corresponds to the remaining $d$-1 features). Finally, the fitness of each feature as a hyper-class is calculated by using the defined Cross Entropy, KL divergence and JS divergence, so as to select the appropriate decision features and construct hyper-class.

The main contributions of this paper are as follows:

- The cross entropy, KL divergence and JS divergence of features in data are defined, and they are used to learn the potential hyper-class information in data.
- The hyper-class representation of data is proposed for the first time, which is different from the hyper-class mentioned in the previous work [11] (*i.e.*, manually annotated class information). The proposed hyper-class representation in this paper uses algorithms to learn the potential class information in features. For example, after the feature of time is represented by hyper-class, there may be only four values in spring, summer, autumn and winter, rather than continuous multiple values.
- we use the collaborative filtering recommendation algorithm to verify the effectiveness of the proposed hyper-class representation, which can improve the speed and accuracy of recommendation.

The rest of this paper is organized as follows. Section 2 briefly reviews previous related work for data representation. Section 3 describes in detail our defined Cross Entropy, KL divergence and JS divergence. Section 4 shows the results of all algorithms on real datasets. Section 5 presents a summary.

## 2. Related work

Data representation is the basis of data science [12,13]. It can help computers to operate data better, summarize data and mine data. In this section, we focus on the data representation model in a database [14] and the data representation with big data [5].

### 2.1. Data representation in database

In database systems, common data representations include relational tables, entity relationship (ER) models and data base task group (DBTG) models [15,16]. They have their own characteristics.

Table 1
The detail of the notations used in this paper.

| | |
|---|---|
| $U$ | dataset |
| $\mathbf{X}_i$ | the $i$-th sample |
| $R$ | the binary relationship on $U$ |
| $|U|$ | the cardinality of set $U$ |
| $E_R(U)$ | the information entropy of the dataset on $R$ |
| $F_k$ | the $k$-th feature |
| $\neg F_k$ | all features except the $k$-th feature. |
| $U/F_k$ | division of data $U$ by feature $F_k$ |
| $U/\neg F_k$ | division of data $U$ by feature $\neg F_k$ |
| $E^{CE}_{F_k,\neg F_k}(U)$ | the cross entropy of the data set on features $F_k$ and $\neg F_k$ |
| $E^{KL}_{F_k,\neg F_k}(U)$ | the KL divergence of dataset $U$ on $F_k$ and $\neg F_k$ |
| $E^{JS}_{F_k,\neg F_k}(U)$ | the JS divergence of dataset $U$ on $F_k$ and $\neg F_k$ |

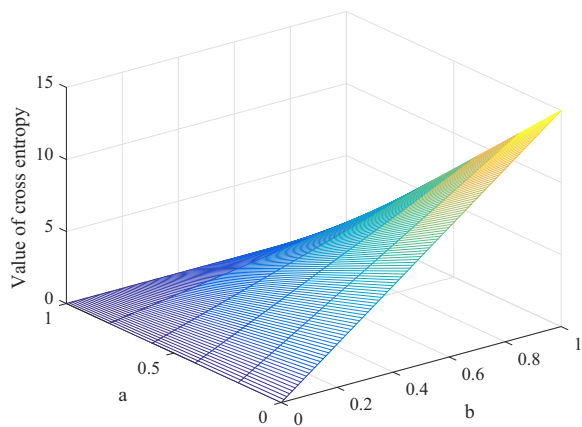

Fig. 2. Cross entropy.

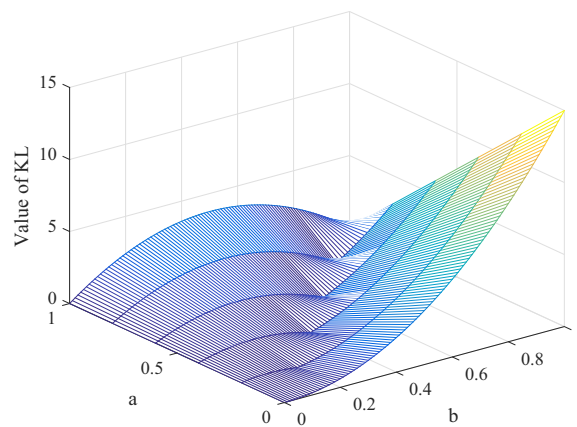

Fig. 3. KL divergence.

Relational table is the theoretical basis of relational model. It is mainly used in relational database [17]. It aims to minimize the occurrence of the same data. It divides all data into multiple tables. One kind of data corresponds to one table, and there are some common values between tables. Sharma et al. used dynamic system global area parameters to adjust the performance of relational databases [18]. Specifically, in order to face the challenges brought by spatial data, they set more than 250 dynamic parameters to indirectly manage the system global area of the running instance of the relational database by adjusting the values of these parameters. In addition, it also analyzes the data related to system parameters and system resources, so as to find the relationship between parameters and resources and improve the performance of the overall system. Mama et al. proposed a flexible relational database, which mainly aims at the fact that customers are not allowed to use fuzzy or imprecise terms in database queries [19]. It points out that the problem of flexible fuzzy database query system can not be solved by using possibility theory and fuzzy logic. Nguyen improves the performance of relational database by selecting the connection algorithm in relational database [20]. Specifically, it first collects baseline query times under different subsets of available join algorithms. Then it uses deep reinforcement learning to select an appropriate subset for each query, which makes it achieve better performance in dense queries. Finally, it found that in the query process, only isolating a "bad" connection and changing its connection algorithm can produce better results. This method shows that reinforcement learning has great potential in the application of relational database. The above data representation based on relational table adopts the idea of linear algebra. The method is simple, easy to understand and widely used, but there will be data redundancy.

ER model is a user oriented representation method, which is widely used in database. It consists of entity, attribute and relationship. Entities are users of data, *e.g.,* human, items *etc*. The characteristics of an entity are attributes, such as a person's height, age, gender and weight. Relationship refers to the relationship between entities. For example, A and B are good friends, and friends are the relationship between them. Lv et al. proposed an entity relation extraction model based on bidirectional maximum entropy [21]. Specifically, it first takes the triple as the entity relationship chain to identify the entity before the relationship and predict its corresponding relationship and entity. Then, it uses two-way long short-term memory and maximum entropy Markov model to extract entities and relationships. Finally, experiments show that it achieves better results than the traditional model. Wu et al. proposed a video summarization method based on entity relationship [22]. Specifically, it uses Wasserstein Gan to solve the problem of low redundancy and instability in Gan. In addition, it also uses the newly proposed patch/ score sum losses to reduce the sensitivity of the model with video length changes. Al Fedaghi proposed a new entity relationship model [23]. Specifically, it extends the ER model by adding attributes to the entity relationship model. At the same time, it also stores structured and time-oriented concepts in it. ER mainly adopts tree structure. It is still often used in the fields of knowledge map, deep learning, natural language processing and so on.

The DBTG model allows a mesh data structure, which consists of data items, data sets, records and set. Data item is the smallest unit of data that can be named. There are two types: numeric type and string type. Data set is a collection of data items. It also has two kinds, namely vector and repetition group. Records are collections of data items and data sets. Set is the most important component of DBTG model. It is a collection of records. Through the concept of set, DBTG can describe the corresponding relationship between any entity without data duplication. Thomas et al. proposed two verification algorithms for pointer values in DBTG database [24]. On the one hand, it uses the "pointer" in modern programming language to judge the exceptions of instances in sets and records. On the other hand, it checks instances in the collection to complete pointer validation. It avoids data duplication. Hawley et al. put forward some shortcomings of DBTG [25]. Specifically, they pointed out the problems in integrity assurance, rapid system response and programmer burden. DBTG adopts a network structure, which can represent diversity relations and semantics. However, there is not much new work on DBTG, which has stayed in the last century.

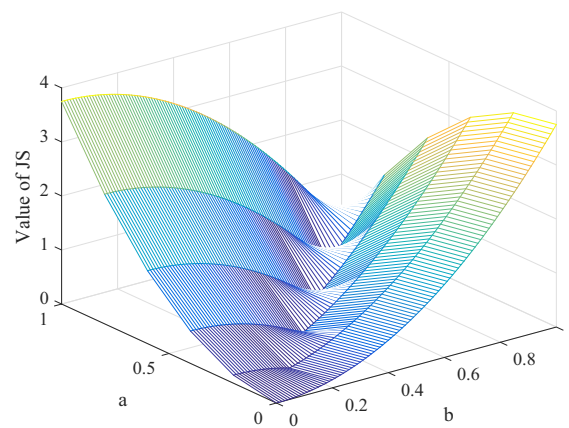

Fig. 4. JS divergence.

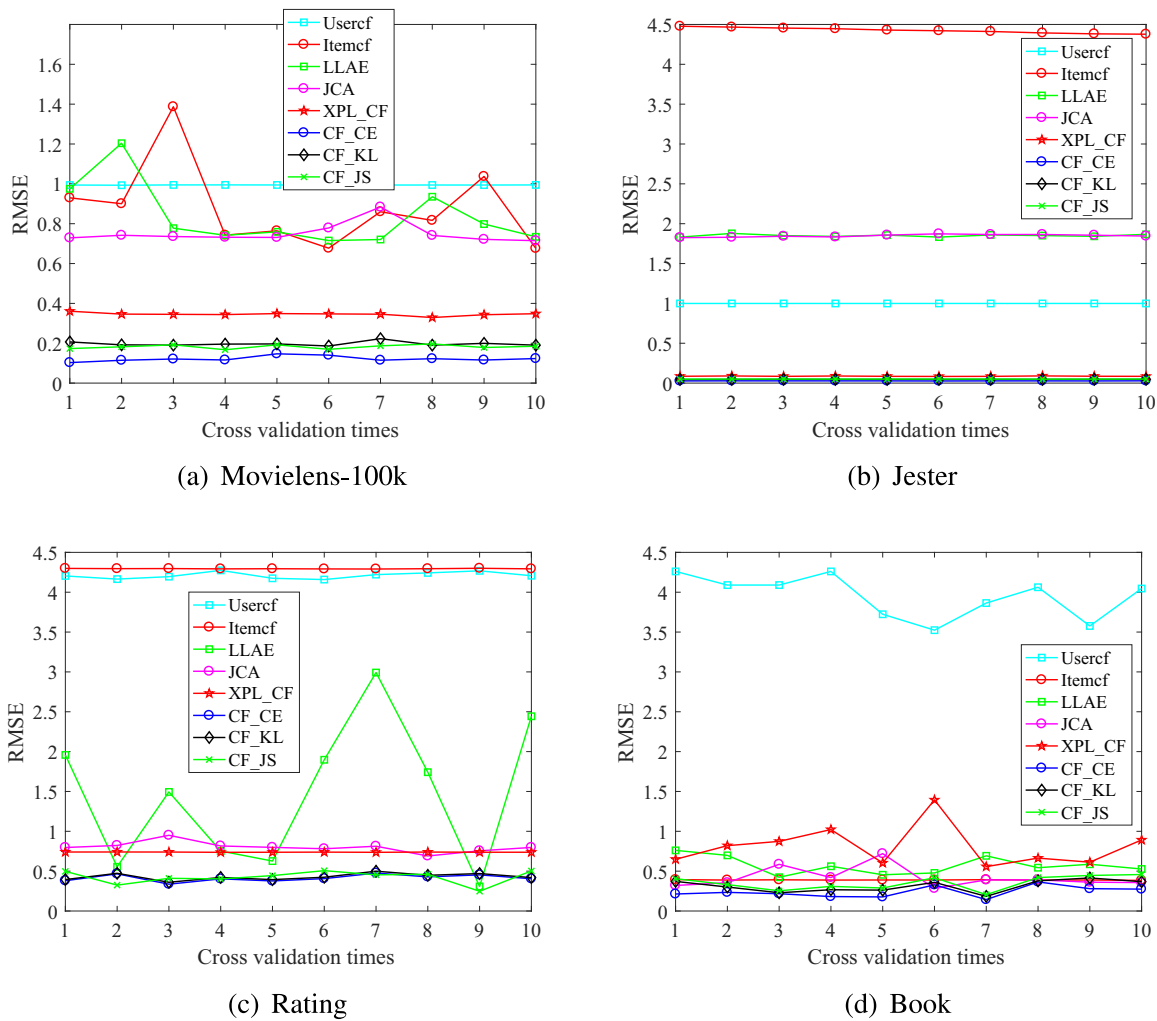

Fig. 5. Running results of all algorithms on RMSE.

Table 2
Average RMSE results of all algorithms on real datasets.

| Datasets | Usercf | Itemcf | LLAE | JCA | XPL_CF | CF_CE | CF_KL | CF_JS |
|---|---|---|---|---|---|---|---|---|
| Movielens-100 k | 0.9942 | 0.8791 | 0.8361 | 0.7508 | 0.3458 | **0.1218** | 0.1971 | 0.1824 |
| Jester | 0.9992 | 4.4262 | 1.8508 | 1.8490 | 0.0863 | **0.0274** | 0.0467 | 0.0501 |
| Rating | 4.2110 | 4.2950 | 1.4770 | 0.8021 | 0.7384 | **0.4132** | 0.4304 | 0.4274 |
| Book | 3.9498 | 0.3894 | 0.5726 | 0.4183 | 0.8087 | **0.2413** | 0.3135 | 0.3534 |
| average value | 2.5386 | 2.4974 | 1.1841 | 0.9551 | 0.4948 | **0.2009** | 0.2469 | 0.2533 |

Table 3
Average MAE results of all algorithms on real datasets.

| Datasets | Usercf | Itemcf | LLAE | JCA | XPL_CF | CF_CE | CF_KL | CF_JS |
|---|---|---|---|---|---|---|---|---|
| Movielens-100 k | 0.9937 | 1.1261 | 1.0292 | 0.7260 | 1.4883 | **0.1594** | 0.2593 | 0.2312 |
| Jester | 0.9992 | 3.1380 | 1.4732 | 1.4719 | 0.2334 | **0.0231** | 0.0476 | 0.0540 |
| Rating | 3.7891 | 4.2815 | 1.1748 | 0.6016 | 0.6798 | **0.3388** | 0.3703 | 0.3996 |
| Book | 3.4418 | **0.0608** | 0.4521 | 0.6089 | 2.1733 | 0.2639 | 0.3607 | 0.4562 |
| average value | 2.3060 | 2.1516 | 1.0323 | 0.8521 | 1.1437 | **0.1963** | 0.2595 | 0.2853 |

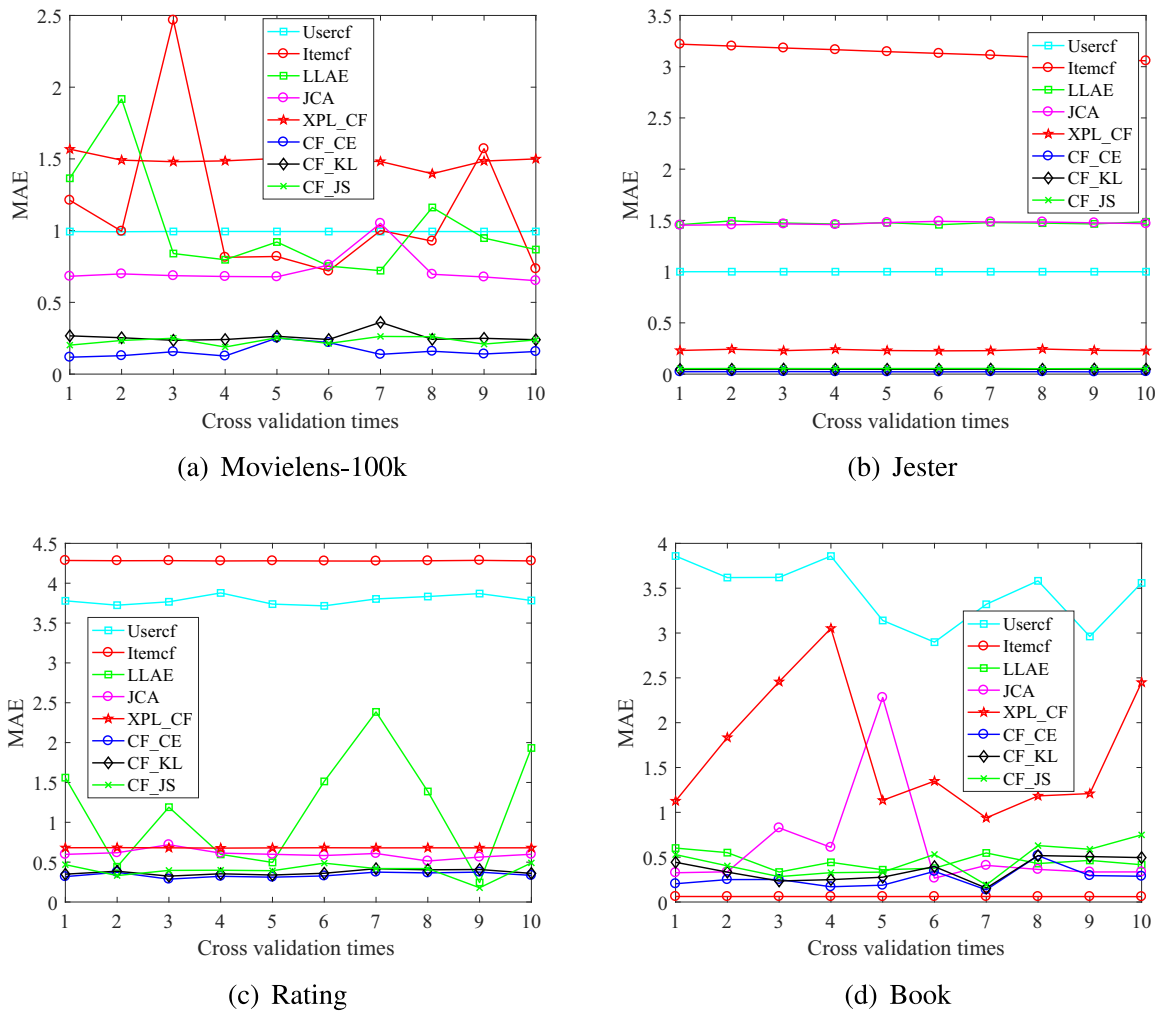

Fig. 6. Running results of all algorithms on MAE.

### 2.2. Data representation in big data

According to whether the data has label information or not, data representation learning can be divided into supervised, unsupervised and semi supervised [26].

In the supervised data representation, the most classic is LDA algorithm. Its core idea is to find a suitable projection direction, so that the data of different classes are as far as possible and the data of the same classes are as close as possible. On this basis, Yang et al. proposed 2D PCA. It directly uses the original image to construct the covariance matrix and its eigenvector for data representation. Zhou et al. proposed a non dominated sorting genetic algorithm to improve supervised data representation [27]. Specifically, it first uses the accuracy first control operator to select individuals with high classification accuracy to survive. Then it designs mutation retry operator and combination operator to make the algorithm converge faster. Finally, it selects the appropriate feature subset from the obtained pareto solution. Lu and Chen proposed a supervised feature selection algorithm by combining feature weight and graph matrix regression [28]. On the one hand, it encodes the correlation between features through the weight matrix. On the other hand, it uses regularization term to adaptively learn graph matrix in low dimensional space. Finally, it uses alternating iterative optimization to obtain the final value of each variable for feature selection. Chen et al. proposed a supervised feature selection algorithm by embedding discrimination into sparse matrix regression [29]. Specifically, it first uses the left–right regression matrix to consider the relationship between different types of data, and then maximizes the average margin of data to obtain nonlinear discriminant embedding. Finally, it can obtain the nonlinear embedding and linear approximation of data by the iterative optimization algorithm.

In unsupervised data representation, PCA algorithm is the classic one. Its key idea is to maximize the variance of each dimension of the original data in the shadow space, in order to remove the redundant dimensions of the original data, so as to obtain the low-dimensional representation of the data. Compared with supervised feature selection, unsupervised feature selection is self-learning of data without data class label. Tang et al. proposed an unsupervised multi-view feature selection algorithm [30]. It mainly considers the application of cross view information in multi-view learning. In addition, it uses the regular term of similarity graph to maintain the local information of data, and uses $l_{2,1}$- norm limits the relationship matrix under each view, so as to select different features from different views. Shang et al. proposed an unsupervised feature selection algorithm based on representation learning [31]. Specifically, it first uses representation learning to characterize the sample structure and feature structure of data. Then it uses the low dimensional potential representation matrix as the label matrix to solve the problem of label missing. Finally, it unifies the representation matrix and transformation matrix of feature space to select features. Miao et al. proposed an unsupervised feature selection algorithm based on graph regularization [32]. It clearly defines a feature selection matrix composed of 0 and 1 to select the appropriate feature subset. In addition, it also takes manifold regularization as an

**Table 4**
The running cost of all algorithms on real datasets (second).

| Datasets | Usercf | Itemcf | LLAE | JCA | XPL_CF | CF_CE | CF_KL | CF_JS |
|---|---|---|---|---|---|---|---|---|
| Movielens-100 k | 40.00 | 1515.24 | 32.82 | 45.19 | 29.91 | 14.74 | 13.60 | **9.97** |
| Jester | 8238.22 | 1597.31 | 1522.64 | 776.23 | 341.47 | 145.79 | **135.49** | 135.69 |
| Rating | 499.14 | 3490.61 | 3490.61 | **146.63** | 567.42 | 208.39 | 212.94 | 207.50 |
| Book | 22.24 | 33.88 | 9.47 | 4.98 | 20.19 | **0.37** | **0.37** | **0.37** |
| average value | 2199.90 | 1659.25 | 1263.89 | 243.26 | 239.75 | 92.32 | 90.60 | **88.38** |

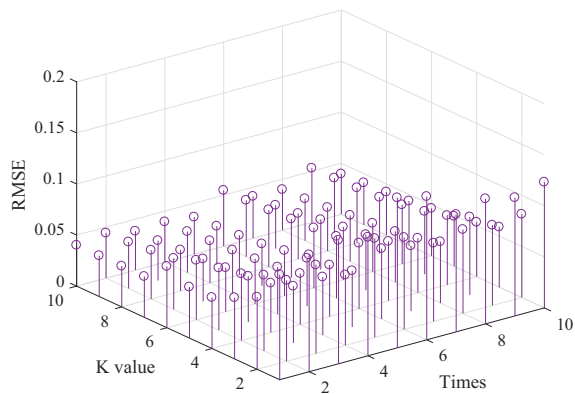

(a) Movielens-100k

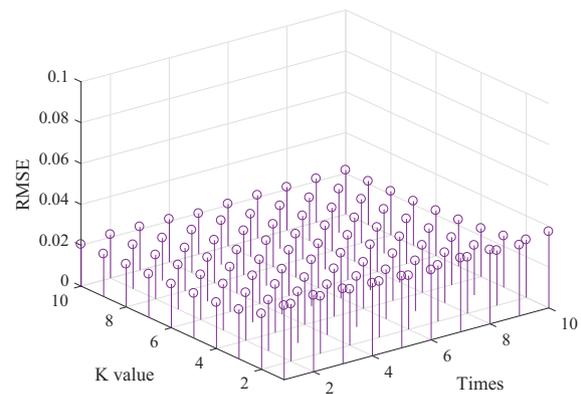

(b) Jester

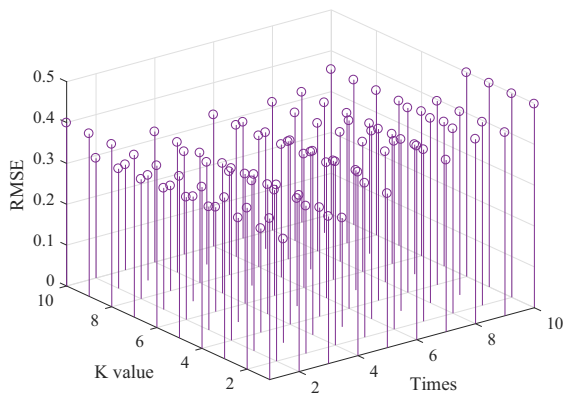

(c) Rating

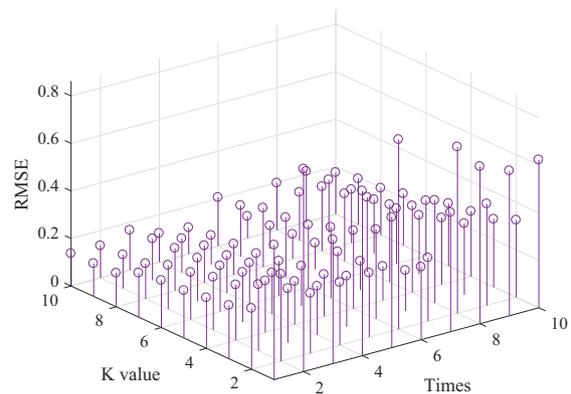

(d) Book

**Fig. 7.** The parameter sensitivity of the proposed CF_CE on the RMSE.

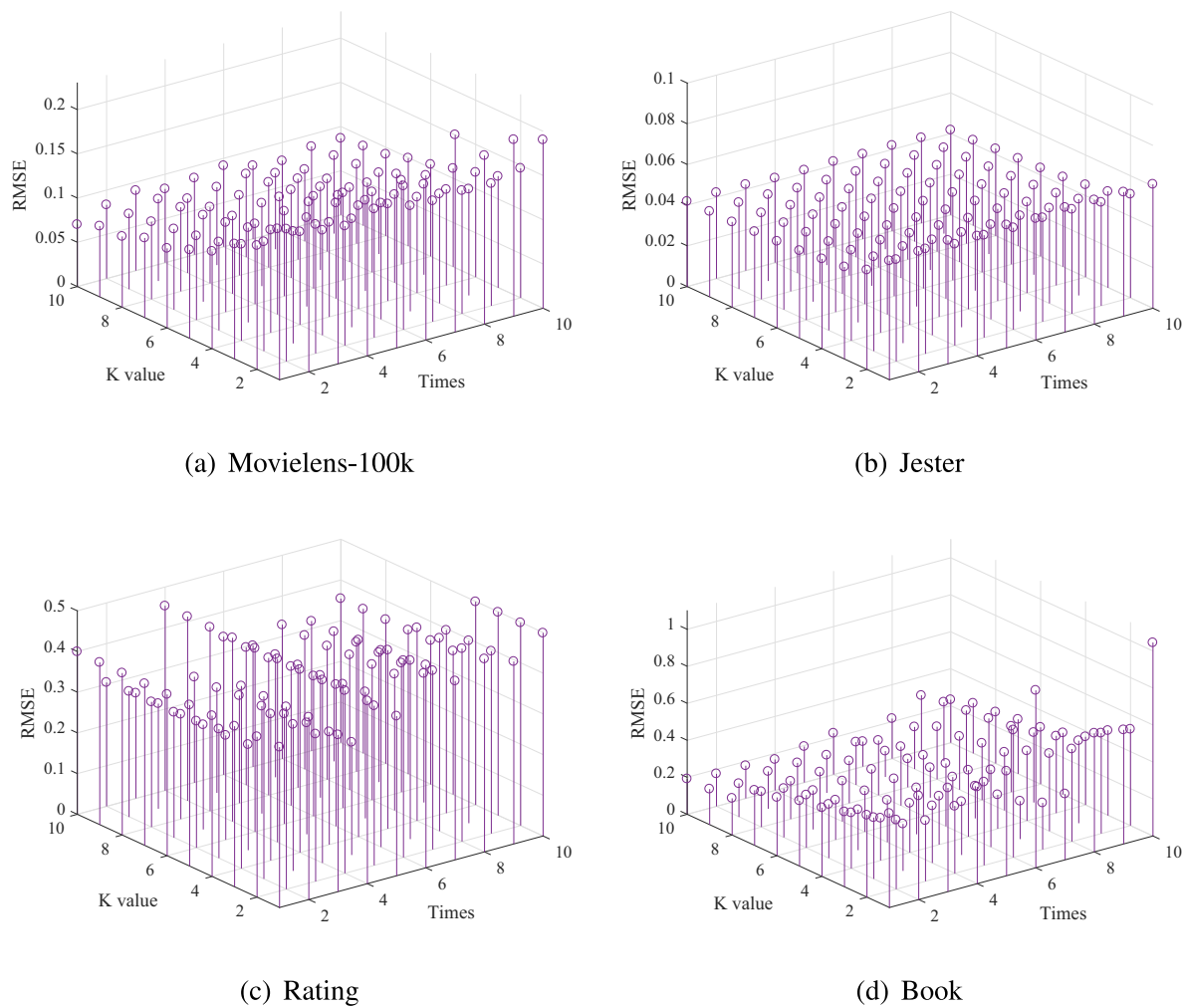

(a) Movielens-100k      (b) Jester

(c) Rating      (d) Book

**Fig. 8.** The parameter sensitivity of the proposed CF_KL on the RMSE.

auxiliary means of local linear embedding to find relevant features. Li et al. proposed an unsupervised low dimensional representation of data [33]. Specifically, it first maps each feature to the kernel space, and each feature corresponds to a kernel matrix. Then it applies a weight to each kernel matrix to learn important features. Finally, it uses $l_1$- norm to obtain the low dimensional representation of data. The method is a nonlinear algorithm.

In real life, data often has only a part of labels. At this time, it is necessary to learn both unlabeled data and labeled data at the same time. This leads to that semi-supervised learning has become a hot research field. Li et al. proposed a semi-supervised feature selection algorithm based on manifold embedding [34]. It uses generalized uncorrelated constraints to solve the problem that the closed solution cannot be obtained in ridge regression. In addition, the embedded manifold structure can better save the topology information of data. Chen et al. proposed a semi-supervised feature selection algorithm based on structural manifold learning [35]. It can learn more structure diagrams than known classes, and use the nearest neighbors of all data to construct submanifold structures in labeled data and unlabeled data. Wang et al. proposed a semi-supervised feature selection algorithm based on sparse discriminant least squares [36]. It uses $l_{2,p}$- norm to calculate the row coefficients of the feature selection matrix, so as to select the appropriate feature subset.

In addition, Zeng et al. proposed a switched particle swarm optimization algorithm based on dynamic neighborhood [37]. It combines differential evolution algorithm and particle swarm optimization algorithm, and uses dynamic distance neighborhood to adjust the best position of individuals, so as to make full use of the population evolution information in the whole population. Luo et al. proposed a new particle swarm optimization algorithm [38]. Specifically, it first adds more dynamic information to particle swarm optimization to avoid premature convergence. Then it uses latent factor analysis to extract useful information from high-dimensional sparse matrix. Finally, it is used to efficiently represent high-dimensional sparse matrices. This method can be regarded as a representation of high-dimensional sparse data. Since this paper uses the most traditional collaborative filtering recommendation algorithm to verify the effectiveness of the proposed hyper-class representation, we briefly introduce collaborative filtering recommendation. The core of collaborative filtering recommendation is to calculate the distance or similarity between target users and known users, and recommend items to target users with the hobbies of known users with high similarity. For example, Yue et al. introduced many current collaborative filtering recommendation algorithms and pointed out their five scenarios in health recommendation [39], namely, diet recommendation, lifestyle recommendation, training recommendation, patient and doctor decision-making, and disease-related prediction.

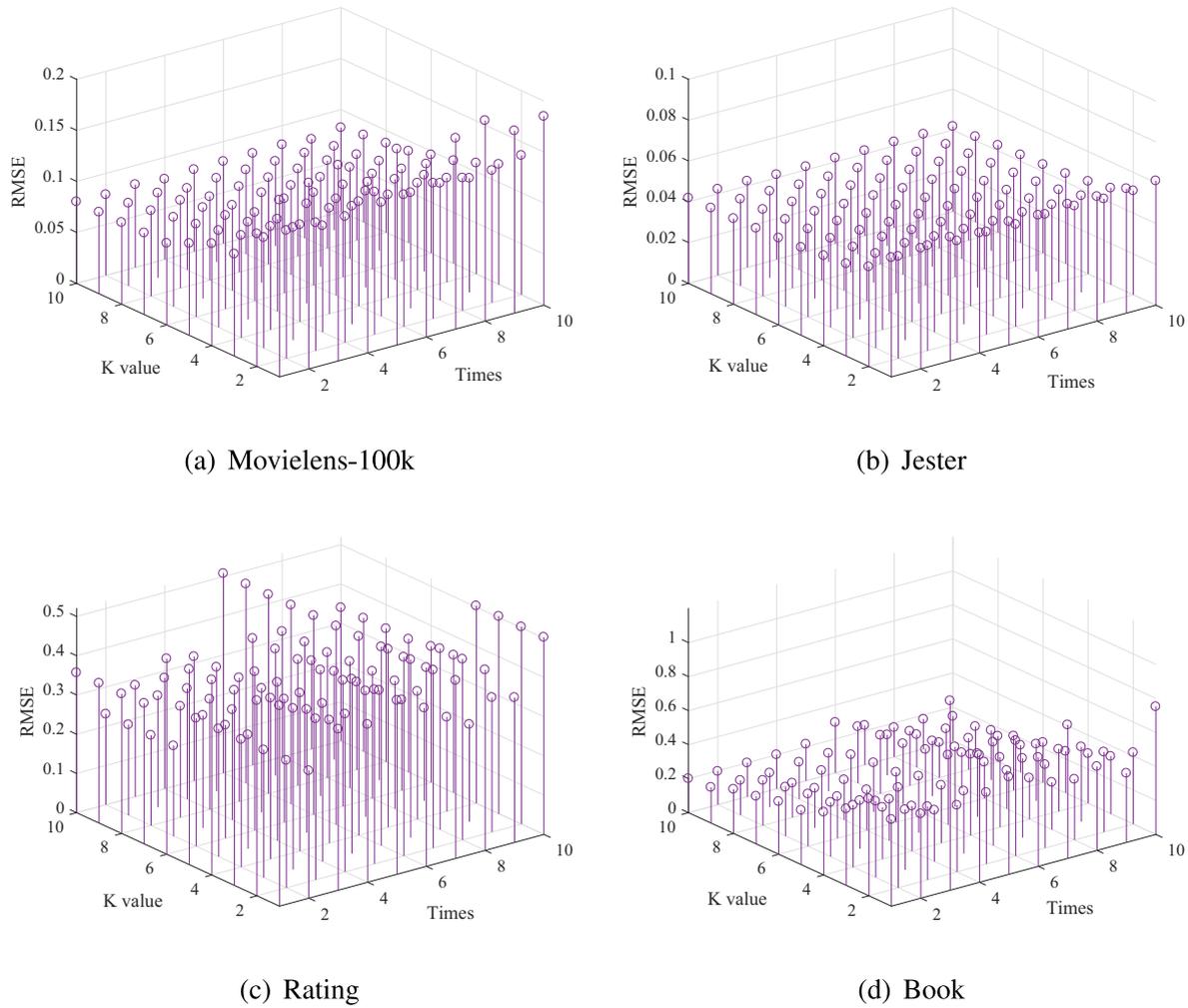

**Fig. 9.** The parameter sensitivity of the proposed CF_JS on the RMSE.

Whether it is the data representation method in the database or the data representation method under big data, they are the basis for the operation of the algorithm. These data representation methods have their own characteristics. For example, the entity relationship model can well describe the relationship of each entity and the characteristics of each entity. But its disadvantage is the same as that of relational tables. After they are applied to the database, the query efficiency is not high and the speed is slow. DBTG model allows mesh data structure, which can represent the association between data. But its disadvantage is that it has a large amount of data duplication, which increases the storage burden. In the big data, common data representation includes low dimensional representation, i.e., it reduces the dimension of high-dimensional data to learn the low dimensional representation of data. For example, LDA is a supervised low dimensional representation technology, while PCA is an unsupervised low dimensional representation technology. Both can learn the low dimensional representation of data. Their disadvantage is that they are not suitable for learning the low dimensional representation of non gaussian data and may cause over fitting. The proposed hyper-class representation in this paper is different from the previous data representation. It can not only reduce the amount of data storage, but also make the potential hyper-class information appear in the data.

## 3. Method

### 3.1. Notation

In this paper, we use $x_i$ to represent the $i$-th sample. In addition, $U$ represents the entire dataset. $R$ represents the division of data, $E_R(U)$ represents the information entropy of the dataset on $R$, $E^{CE}_{F_k,\neg F_k}(U)$ represents the cross entropy of the dataset on features $F_k$ and $\neg F_k$. $E^{KL}_{F_k,\neg F_k}(U)$ represents the KL divergence of dataset $U$ on $F_k$ and $\neg F_k$, $F_k$ represents the $k$-th feature, and $\neg F_k$ represents all features except the $k$-th feature.

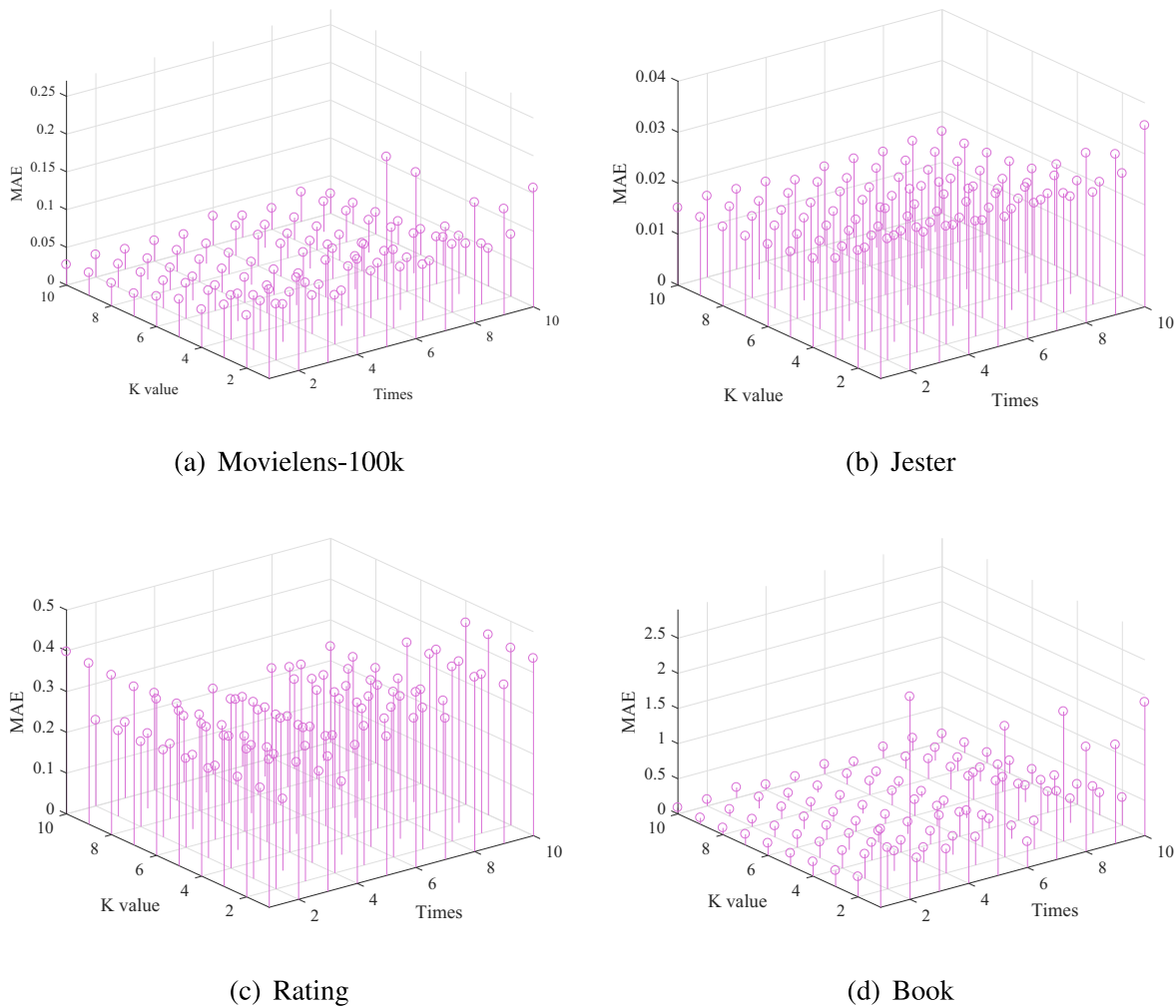

(a) Movielens-100k  (b) Jester

(c) Rating  (d) Book

**Fig. 10.** The parameter sensitivity of the proposed CF_CE on the MAE.

We summarize these notations used in our paper in Table 1.

### 3.2. Motivation

Existing data representation is usually based on attribute values, i.e., given a dataset $X \in \mathbb{R}^{n \times d}$, it means that there are $n$ samples, and each sample has $d$ attributes. On this basis, the process of processing data becomes attribute-centered calculation. In unsupervised learning, this form not only consumes memory, but also has no label information. Although the low dimensional representation of data can reduce the storage capacity of data, it still can not show the potential class information in the data. Therefore, this paper proposes a hyper-class representation, which can not only reduce the storage of attribute values, but also store some appropriate continuous attribute values into discrete attribute values. It can also make the potential class information in the data appear. This is extremely useful for subsequent data mining algorithms. For example, in the recommendation algorithm, given a target user, most of the known users can be excluded in advance according to the hyper-class information of the data, so as to find similar users of the target user quickly and accurately.

### 3.3. Information entropy

The concept of entropy is the core of all kinds of entropy calculation. The reasons are as follows: 1. The success of various entropy calculations largely depends on the entropy of the internal chaos of the system. 2. It has important applications in cybernetics, probability theory, number theory, astrophysics, life science and other fields. More specific definitions in different disciplines can promote the development of human society.

**Definition 1** [40]. Give an unlabeled data set $S = (U, R)$, where $U$ is a non empty finite data set, $R$ is the binary relationship on $U$, and for $\forall x_i \in U$, the R relationship set on $x_i$ is:

$$R(x_i) = \{x_j \in U | (x_i, x_j) \in R\} \quad (1)$$

**Definition 2** [41]. Given an unlabeled data set $S = (U, R)$, the information entropy of the data set on $R$ is:

$$E_R(U) = \sum_{i=1}^{n} \frac{|R(x_i)|}{|U|}(1 - \frac{|R(x_i)|}{|U|}) \quad (2)$$

As can be seen from Eq. (2), $0 \leqslant E_R(U) \leqslant \frac{n}{4}$ is always true. The information entropy on R clearly reveals the distinguishability of data. When $0 < \frac{|R(x_i)|}{|U|} \leqslant \frac{1}{2}$, $E_R(U)$ is monotonically increasing. At this time, R contains fewer ordered pairs, which shows that most data in $U$ can be easily distinguished. The smaller $|R(x_i)|$, the smaller $E_R(U)$. When $\frac{1}{2} < \frac{|R(x_i)|}{|U|} \leqslant 1$, the smaller $|R(x_i)|$, the larger $E_R(U)$.

### 3.4. Cross entropy

In rough set theory, the defined information entropy is based on the sample relationship, which does not take into account the relationship between features. Moreover, information entropy can not calculate the relationship between the two divisions, so it can not select the decision features. For this reason, we define the cross entropy of the feature, as shown below.

**Definition 3.** Given an unlabeled data set $S = (U, F_k \cup \neg F_k)$, where $F$ is the feature set, $U/F_k = \{\alpha_1, \alpha_2, \ldots, \alpha_n\}, U/\neg F_k = \{\beta_1, \beta_2, \ldots, \beta_m\}$. The cross entropy of dataset $U$ on $F_k$ and $\neg F_k$ is:

$$E^{CE}_{F_k,\neg F_k}(U) = \sum_{i=1}^{n}\sum_{j=1}^{m}\frac{|\alpha_i|}{|U|}(1 - \frac{|\beta_j|}{|U|}) \quad (3)$$

As shown in Eq. (3), cross entropy can be used to calculate the difference between two divisions $U/F_k$ and $U/\neg F_k$. The smaller the value of $E^{CE}_{F_k,\neg F_k}(U)$, it indicates that the partition $U/F_k$ generated by feature $F_k$ is more distinguishable and can be defined as a hyper-class. As shown in Fig. 2, we assume that feature $F_k$ divides the data into 5 classes and $\neg F_k$ divides the data into 3 classes, i.e., $n = 5$ and $m = 3$. At this time, the value of $E^{CE}_{F_k,\neg F_k}(U)$ increases with the increase of the value of $\frac{|\alpha_i|}{|U|}$. The smaller the $E^{CE}_{F_k,\neg F_k}(U)$ value, the smaller the difference between $U/F_k$ and $U/\neg F_k$. The greater the degree to which $F_k$ can be used as a decision feature.

The cross entropy of the data set on $F_k$ and $\neg F_k$ has the following properties:
**Property 1**: nonnegativity: $0 \leqslant E^{CE}_{F_k,\neg F_k}(U) \leqslant nm$.
**Property 2**: asymmetry.

**Proof.** Let $b = \frac{|\alpha_i|}{|U|}$ and $a = \frac{|\beta_j|}{|U|}$. It ia clear that $0 \leqslant a, b \leqslant 1$. When we assume that the values of $n$ and $m$ are both 1, we can get the following formula:

$$\frac{E^{CE}_{F_k,\neg F_k}(U)}{\partial a} = -b \quad (4)$$

$$\frac{E^{CE}_{F_k,\neg F_k}(U)}{\partial b} = 1 - a \quad (5)$$

It is easy to obtain the minimum value of 0 for $E^{CE}_{R,V}(U)$ when $b = 0$. When $a = 0, b = 1, E^{CE}_{R,V}(U)$ achieves the maximum value of 1. As shown in Fig. 2.

**Proof.** To prove asymmetry, we only need to prove $E^{CE}_{F_k,\neg F_k}(U) - E^{CE}_{\neg F_k,F_k}(U) \neq 0$, as shown in the following formula:

$$\begin{aligned}&E^{CE}_{F_k,\neg F_k}(U) - E^{CE}_{\neg F_k,F_k}(U) \\ &= \sum_{i=1}^{n}\sum_{j=1}^{m}\frac{|\alpha_i|}{|U|}(1 - \frac{|\beta_j|}{|U|}) \\ &\quad -(\sum_{j=1}^{m}\sum_{i=1}^{n}\frac{|\beta_j|}{|U|}(1 - \frac{|\alpha_i|}{|U|})) \\ &= \sum_{i=1}^{n}\sum_{j=1}^{m}\frac{|\alpha_i|}{|U|} - \frac{|\alpha_i||\beta_j|}{|U|^2} - \frac{|\beta_j|}{|U|} + \frac{|\beta_j||\alpha_i|}{|U|^2} \\ &= \sum_{i=1}^{n}\sum_{j=1}^{m}\frac{|\alpha_i|}{|U|} - \frac{|\beta_j|}{|U|} \\ &\neq 0\end{aligned} \quad (6)$$

### 3.5. KL divergence

Cross entropy can measure the difference between two partitions to a certain extent, and it can be regarded as a special case of KL divergence. Further, we define KL divergence based on cross entropy.

**Definition 4.** Given an unlabeled data set $S = (U, F_k \cup \neg F_k)$, the relative entropy or KL divergence of the data set on $F_k$ and $\neg F_k$ is:

$$E^{KL}_{F_k,\neg F_k}(U) = \sum_{i=1}^{n}\sum_{j=1}^{m}\left|\frac{|\alpha_i|^2}{|U|^2} - \frac{|\alpha_i||\beta_j|}{|U|^2}\right| \quad (7)$$

As shown in Fig. 3, the value of KL divergence changes when $n = 5$ and $m = 3$. As can be seen from Fig. 3, it shows a three-stage change with the increase of $\frac{|\alpha_i|}{|U|}$ value, i.e., first increasing, then decreasing, and then increasing. It is different from cross entropy. It takes $\frac{|\beta_j|}{|U|} = \frac{|\alpha_i|}{|U|}$ as the dividing line. The reason is that we added the absolute value symbol to ensure that it is non negative. Similarly, the smaller the value of $E^{KL}_{F_k,\neg F_k}(U)$, the greater the degree to which $F_k$ can be used as a decision feature.

The relative entropy or KL divergence of the data set on $F_k$ and $\neg F_k$ has the following properties:
**Property 1**: nonnegativity: $E^{KL}_{F_k,\neg F_k}(U) \geqslant 0$, if and only if $U/F_k = U/\neg F_k, E^{KL}_{F_k,\neg F_k}(U) = 0$.
**Property 2**: asymmetry: $E^{KL}_{F_k,\neg F_k}(U) \neq E^{KL}_{\neg F_k,F_k}(U)$.
**Property 3**: $E^{KL}_{F_k,\neg F_k}(U) = \left|E^{CE}_{F_k,\neg F_k}(U) - E_{F_k}(U)\right|$.

**Proof.** obviously, because $E^{KL}_{F_k,\neg F_k}(U)$ takes the absolute value of time, $E^{KL}_{F_k,\neg F_k}(U) \geqslant 0$ must be true. When $U/F_k = U/\neg F_k$, the following formula holds:

$$\begin{aligned}E^{KL}_{F_k,\neg F_k}(U) &= \sum_{i=1}^{n}\sum_{j=1}^{m}\left|\frac{|\alpha_i|^2}{|U|^2} - \frac{|\alpha_i||\beta_j|}{|U|^2}\right| \\ &= \sum_{i=1}^{n}\sum_{i=1}^{n}\left|\frac{|\alpha_i|^2}{|U|^2} - \frac{|\alpha_i||\alpha_i|}{|U|^2}\right| \\ &= \sum_{j=1}^{m}\sum_{j=1}^{m}\left|\frac{|\beta_j|^2}{|U|^2} - \frac{|\beta_j||\beta_j|}{|U|^2}\right| \\ &= 0\end{aligned} \quad (8)$$

**Proof.** To prove asymmetry, we only need to prove $E^{KL}_{F_k,\neg F_k}(U) - E^{KL}_{\neg F_k,F_k}(U) \neq 0$. We discuss it in the following two cases respectively.

(1) When $|\alpha_i| > |\beta_j|$, then

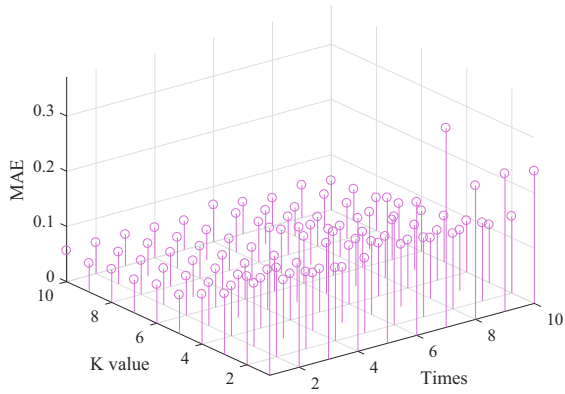

(a) Movielens-100k

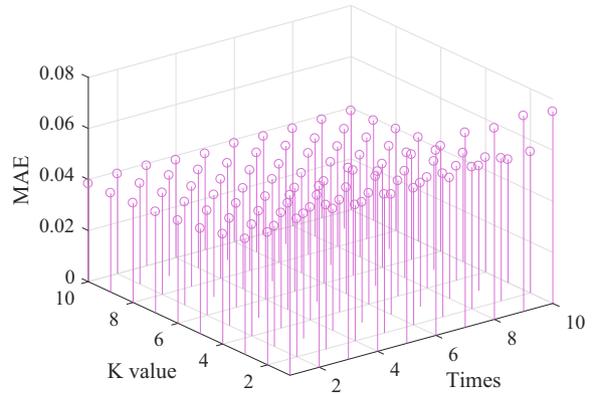

(b) Jester

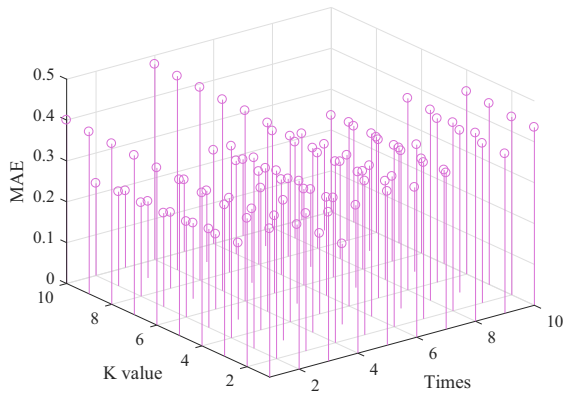

(c) Rating

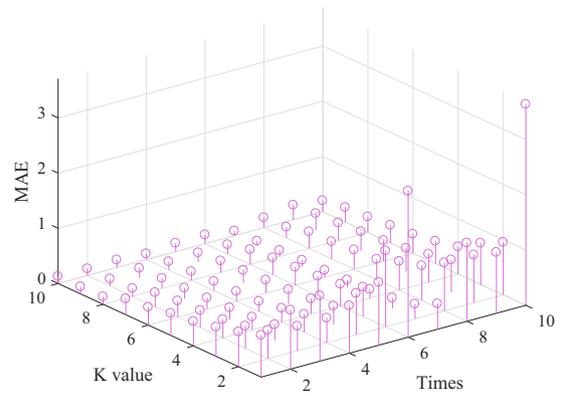

(d) Book

**Fig. 11.** The parameter sensitivity of the proposed CF_KL on the MAE.

$$E_{F_k,\neg F_k}^{KL}(U) - E_{\neg F_k,F_k}^{KL}(U)$$

$$= \sum_{i=1}^{n}\sum_{j=1}^{m}\left|\frac{|\alpha_i|^2}{|U|^2} - \frac{|\alpha_i||\beta_j|}{|U|^2}\right|$$

$$-\sum_{j=1}^{m}\sum_{i=1}^{n}\left|\frac{|\beta_j|^2}{|U|^2} - \frac{|\beta_j||\alpha_i|}{|U|^2}\right|$$

$$= \sum_{i=1}^{n}\sum_{j=1}^{m}\frac{|\alpha_i|^2}{|U|^2} - \frac{|\alpha_i||\beta_j|}{|U|^2}$$

$$-\sum_{j=1}^{m}\sum_{i=1}^{n}\frac{|\beta_j||\alpha_i|}{|U|^2} + \frac{|\beta_j|^2}{|U|^2}$$

$$= \sum_{i=1}^{n}\sum_{j=1}^{m}\frac{|\alpha_i|^2}{|U|^2} - \frac{2|\alpha_i||\beta_j|}{|U|^2} + \frac{|\beta_j|^2}{|U|^2}$$

$$\neq 0$$

(9)

(2) When $|\alpha_i| < |\beta_j|$,

$$E_{F_k,\neg F_k}^{KL}(U) - E_{\neg F_k,F_k}^{KL}(U)$$

$$= \sum_{i=1}^{n}\sum_{j=1}^{m}\left|\frac{|\alpha_i|^2}{|U|^2} - \frac{|\alpha_i||\beta_j|}{|U|^2}\right|$$

$$-\sum_{j=1}^{m}\sum_{i=1}^{n}\left|\frac{|\beta_j|^2}{|U|^2} - \frac{|\beta_j||\alpha_i|}{|U|^2}\right|$$

$$= \sum_{i=1}^{n}\sum_{j=1}^{m}\frac{|\alpha_i||\beta_j|}{|U|^2} - \frac{|\alpha_i|^2}{|U|^2}$$

$$+\sum_{j=1}^{m}\sum_{i=1}^{n}\frac{|\beta_j||\alpha_i|}{|U|^2} - \frac{|\beta_j|^2}{|U|^2}$$

$$= \sum_{i=1}^{n}\sum_{j=1}^{m}\frac{2|\alpha_i||\beta_j|}{|U|^2} - \frac{|\alpha_i|^2}{|U|^2} - \frac{|\beta_j|^2}{|U|^2}$$

$$\neq 0$$

(10)

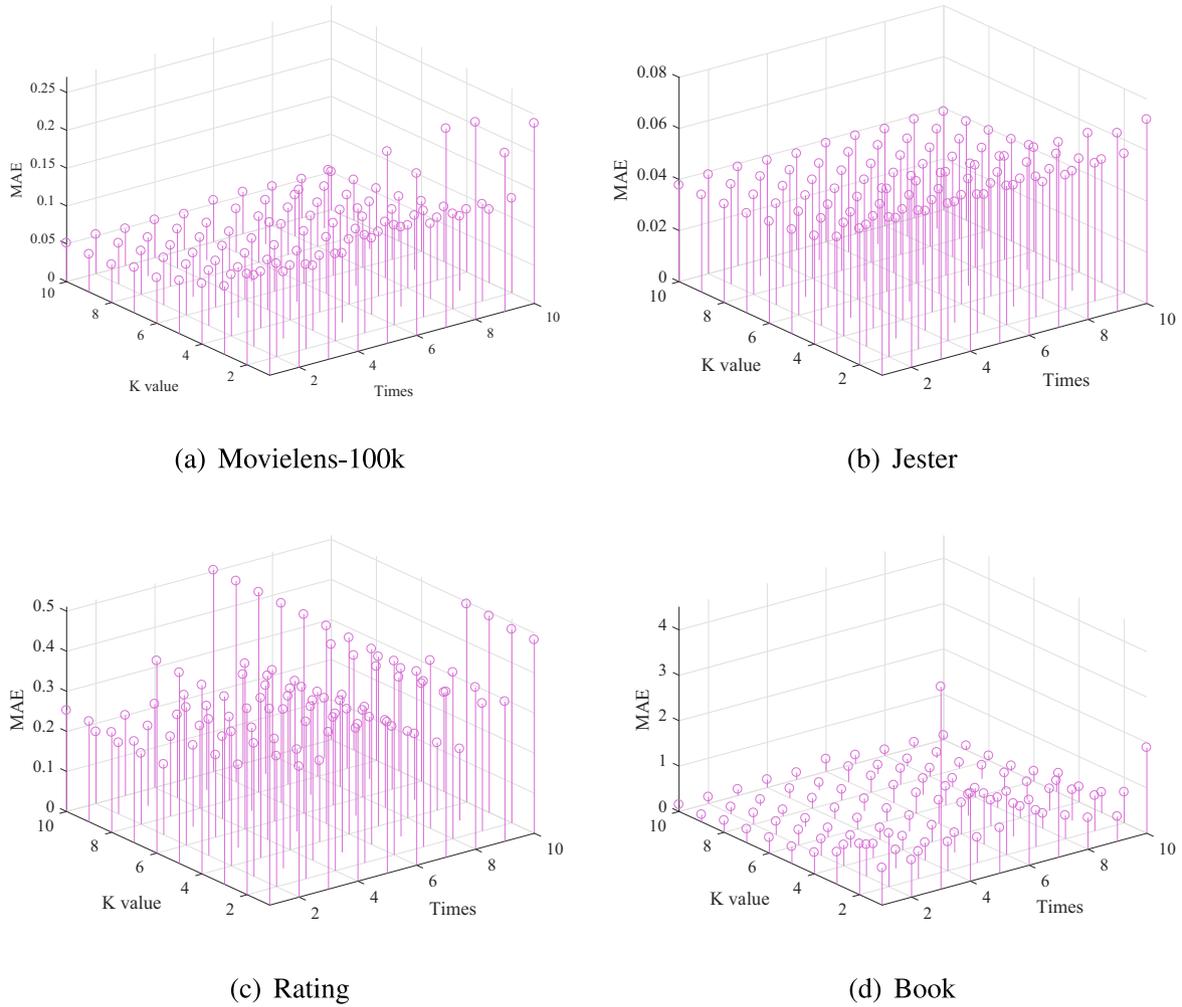

(a) Movielens-100k      (b) Jester

(c) Rating      (d) Book

**Fig. 12.** The parameter sensitivity of the proposed CF_JS on the MAE.

**Proof.** KL divergence is equal to the absolute value of cross entropy minus information entropy. According to Eqs. (2) and (3), we can get:

$$\left| E^{CE}_{F_k, \neg F_k}(U) - E_{F_k}(U) \right|$$
$$= \left| \sum_{i=1}^{n} \sum_{j=1}^{m} \frac{|\alpha_i|}{|U|}(1 - \frac{|\beta_j|}{|U|}) - \sum_{i=1}^{n} \frac{|\alpha_i|}{|U|}(1 - \frac{|\alpha_i|}{|U|}) \right|$$
$$= \left| \sum_{i=1}^{n} \sum_{j=1}^{m} \frac{|\alpha_i|}{|U|} - \frac{|\alpha_i||\beta_j|}{|U|^2} - \frac{|\alpha_i|}{|U|} + \frac{|\alpha_i|^2}{|U|^2} \right| \quad (11)$$
$$= \left| \sum_{i=1}^{n} \sum_{j=1}^{m} \frac{|\alpha_i|^2}{|U|^2} - \frac{|\alpha_i||\beta_j|}{|U|^2} \right|$$
$$= E^{KL}_{F_k, \neg F_k}(U)$$

### 3.6. JS divergence

KL divergence and cross entropy defined by us can measure the difference between the two partitions, and the smaller the difference, the more feature $F_k$ can be used as a decision feature for hyper-class representation. The difference between them is that the information entropy of feature $F_k$ is considered in KL divergence. We can see that KL divergence and cross entropy are equivalent under specific conditions. In addition, we can find that KL divergence and cross entropy do not have symmetry. Therefore, we further define JS divergence.

**Definition 5.** Definition 5: given an unlabeled data set $S = (U, F_k \cup \neg F_k)$, the JS divergence of the data set on $F_k$ and $\neg F_k$ is:

$$E^{JS}_{F_k, \neg F_k}(U) = \tfrac{1}{2} E^{KL}_{F_k, F_M}(U) + \tfrac{1}{2} E^{KL}_{\neg F_k, F_M}(U) \quad (12)$$

where $F_M = \frac{F_k + \neg F_k}{2}$. As shown in Fig. 4, it shows the value change of JS divergence when $n = 5$ and $m = 3$. As can be seen from Fig. 4, JS divergence has symmetry, and the axis of symmetry is $\frac{|\beta_j|}{|U|} = \frac{|\alpha_i|}{|U|}$. JS divergence is based on KL divergence. Therefore, they have the same function. *e.g.*, they can measure the difference between two divisions.

The JS divergence of the dataset on $F_k$ and $\neg F_k$ has the following properties:

**Property 1**: nonnegativity: $E^{JS}_{F_k, \neg F_k}(U) \geqslant 0, E^{JS}_{F_k, \neg F_k}(U) = 0$ if and only if $U/F_k = U/\neg F_k$.

**Property 2**: symmetry: $E^{JS}_{F_k, \neg F_k}(U) = E^{JS}_{\neg F_k, F_k}(U)$.

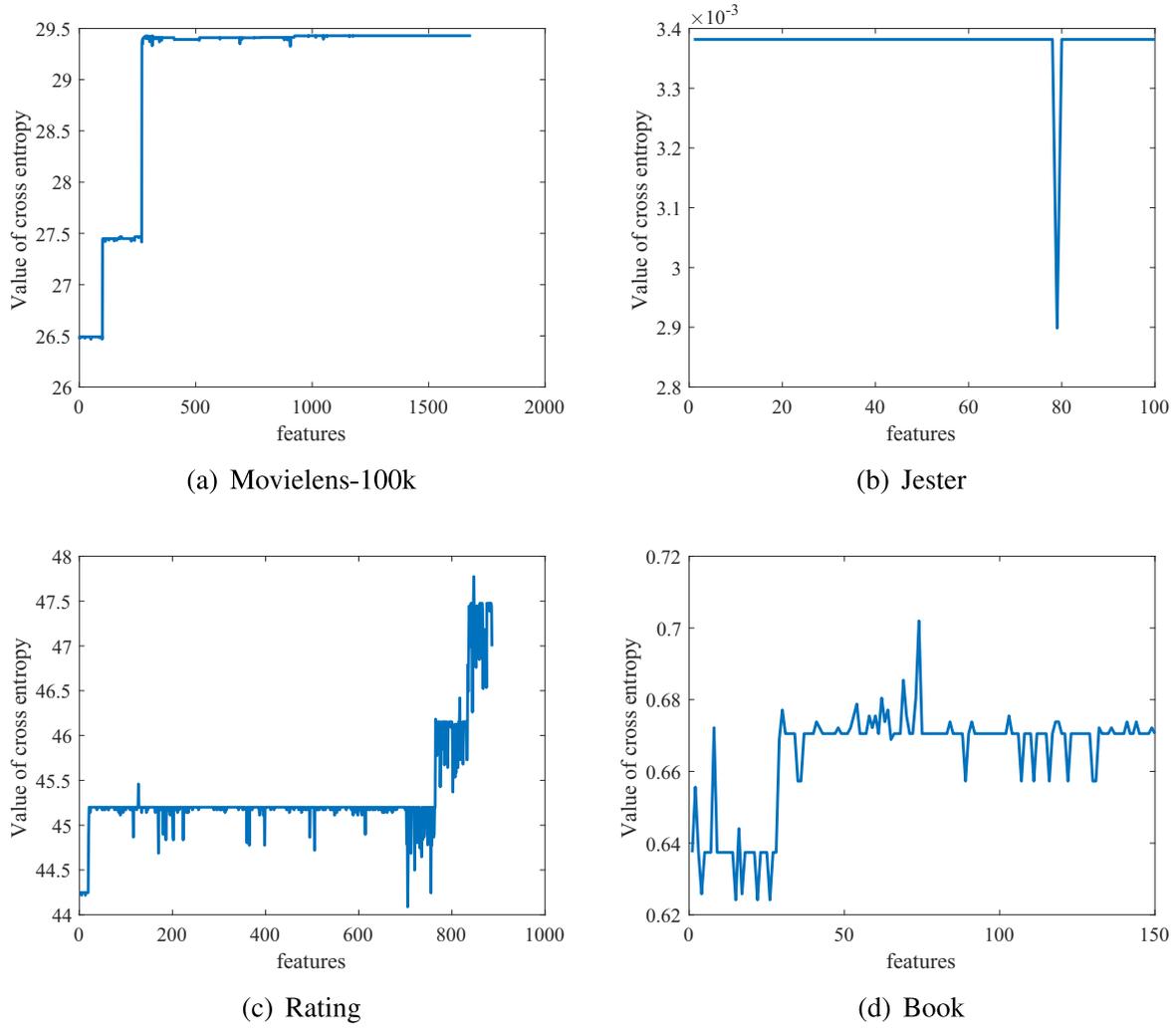

(a) Movielens-100k

(b) Jester

(c) Rating

(d) Book

**Fig. 13.** The cross entropy of each feature on all datasets.

**Proof.** Obviously, JS divergence is based on KL divergence, so it must satisfy nonnegativity. When $U/F_k = U/\neg F_k$, the following formula holds:

$$E^{JS}_{F_k,\neg F_k}(U) = \frac{1}{2}\sum_{i=1}^{n}\sum_{j=1}^{m}\left|\frac{|\alpha_i|^2}{|U|^2} - \frac{|\alpha_i|(|\alpha_i|+|\beta_j|)}{2|U|^2}\right|$$

$$+ \frac{1}{2}\sum_{j=1}^{m}\sum_{i=1}^{n}\left|\frac{|\beta_i|^2}{|U|^2} - \frac{|\beta_i|(|\beta_j|+|\alpha_i|)}{2|U|^2}\right|$$

$$= \frac{1}{2}\sum_{i=1}^{n}\sum_{i=1}^{n}\left|\frac{|\alpha_i|^2}{|U|^2} - \frac{|\alpha_i|(|\alpha_i|+|\alpha_i|)}{2|U|^2}\right|$$

$$+ \frac{1}{2}\sum_{i=1}^{n}\sum_{i=1}^{n}\left|\frac{|\alpha_i|^2}{|U|^2} - \frac{|\alpha_i|(|\alpha_i|+|\alpha_i|)}{2|U|^2}\right|$$

$$= \frac{1}{2}\sum_{j=1}^{m}\sum_{j=1}^{m}\left|\frac{|\beta_j|^2}{|U|^2} - \frac{|\beta_j|(|\beta_j|+|\beta_j|)}{2|U|^2}\right|$$

$$+ \frac{1}{2}\sum_{j=1}^{m}\sum_{j=1}^{m}\left|\frac{|\beta_j|^2}{|U|^2} - \frac{|\beta_j|(|\beta_j|+|\beta_j|)}{2|U|^2}\right|$$

$$= 0$$

(13)

**Proof.** To prove symmetry, we only need to prove $E^{JS}_{F_k,\neg F_k}(U) - E^{JS}_{\neg F_k,F_k}(U) = 0$, and the following formula holds:

$$E^{JS}_{F_k,\neg F_k}(U) - E^{JS}_{\neg F_k,F_k}(U)$$

$$= \frac{1}{2}\sum_{i=1}^{n}\sum_{j=1}^{m}\left|\frac{|\alpha_i|^2}{|U|^2} - \frac{|\alpha_i|(|\alpha_i|+|\beta_j|)}{2|U|^2}\right|$$

$$+ \frac{1}{2}\sum_{j=1}^{m}\sum_{i=1}^{n}\left|\frac{|\beta_i|^2}{|U|^2} - \frac{|\beta_i|(|\beta_j|+|\alpha_i|)}{2|U|^2}\right|$$

(14)

$$- \frac{1}{2}\sum_{j=1}^{m}\sum_{i=1}^{n}\left|\frac{|\beta_j|^2}{|U|^2} - \frac{|\beta_j|(|\beta_j|+|\alpha_i|)}{2|U|^2}\right|$$

$$- \frac{1}{2}\sum_{i=1}^{n}\sum_{j=1}^{m}\left|\frac{|\alpha_i|^2}{|U|^2} - \frac{|\alpha_i|(|\alpha_i|+|\beta_j|)}{2|U|^2}\right|$$

$$= 0$$

In order to clearly show the flow of the proposed algorithm, we write its pseudo code in Algorithm 1. As shown in Algorithm 1, assuming that the data has a total of $d$ features, we use the defined cross entropy, KL divergence and JS divergence to calculate the suitability of each feature as a decision feature. The smaller the value of cross entropy, KL divergence or JS divergence corresponding to the feature, it indicates that the feature is more likely to be used as a decision feature. Then we use the selected features to construct the hyper-class representation of data. In other words, the process of establishing the hyper-class representation of data is divided into two steps: 1. The potential decision features are selected according to the defined cross entropy, KL divergence and JS divergence. It is different from unsupervised feature selection. Unsupervised feature selection select the feature subset that can better represent the overall information of the data. And we select the features that can be used to build hyper-class. This step is very important. 2. Construct the hyper-class representation of data according to the selected features

.

## 4. Experiments

In order to verify the effectiveness of the proposed hyper-class representation, we first establish a hyper-class for the recommended dataset, and then use the traditional collaborative filtering recommendation algorithm to recommend items for target users according to the established hyper-class. Other comparison algorithms are tested on datasets without hyper-class. Finally, their performances are compared according to the recommended accuracy and speed.

### 4.1. Data sets and comparison algorithms

We downloaded four recommended datasets on the website, *i.e.,* Movielens-100 k [1], Jester [2], Rating [3] and Book [4]. The

**Algorithm 1:** Pseudo code for proposed method.

**Input**: Data set $U, T$;
**Output**: Decision feature and hyper-class;
1  **for** $t = 1 \to d$ **do**
2      **switch** $T$ **do**
3          **case** *CF_CE*
4              Compute $E^{CE}_{F_t, \neg F_t}(U)$ according to formula Eq. (3);
5          **case** *CF_KL*
6              Compute $E^{KL}_{F_t, \neg F_t}(U)$ according to formula Eq. (7);
7          **case** *CF_JS*
8              Compute $E^{JS}_{F_t, \neg F_t}(U)$ according to formula Eq. (12);
9
10          **otherwise**
11              Please enter $T$ equals CF_CE, CF_KL or CF_JS, one of them.;
12          **endsw**
13      **endsw**
14  **end**
15  **if** $T == CF\_CE$ **then**
16      Select the feature corresponding to the minimum value as the decision feature according to $\min\{E^{CE}_{F_0, \neg F_0}(U), E^{CE}_{F_1, \neg F_1}(U), \ldots, E^{CE}_{F_d, \neg F_d}(U)\}$ ;
17  **else if** $T == CF\_KL$ **then**
18      Select the feature corresponding to the minimum value as the decision feature according to $\min\{E^{KL}_{F_0, \neg F_0}(U), E^{KL}_{F_1, \neg F_1}(U), \ldots, E^{KL}_{F_d, \neg F_d}(U)\}$ ;
19  **end**
20  **else if** $T == CF\_JS$ **then**
21      Select the feature corresponding to the minimum value as the decision feature according to $\min\{E^{JS}_{F_0, \neg F_0}(U), E^{JS}_{F_1, \neg F_1}(U), \ldots, E^{JS}_{F_d, \neg F_d}(U)\}$ ;
22  **end**
23  The corresponding division of feature is obtained according to reference [38],*i.e.,* hyper-class ;

---

[1] https://grouplens.org/datasets/movielens/.
[2] https://goldberg.berkeley.edu/jester-data/.
[3] https://www.kaggle.com/skillsmuggler/amazon-ratings.
[4] http://www2.informatik.uni-freiburg.de/ cziegler/BX/.

S. Zhang, J. Li, W. Zhang et al.                                                                                           Neurocomputing 503 (2022) 200–218movielens-100 k dataset was collected by the University of Minnesota through the movielens website. It consists of 943 users scoring 1682 films, of which each user scores at least 20 films. Jester data were collected from April 1999 to May 2003. It included 73421 users' ratings of 100 jokes. The score value is - 10 to 10. The higher the value, the higher the evaluation of the joke. In this paper, we use its subset, *i.e.,* rating data of 100 jokes by 24983 users. The ratings dataset is the ratings of beauty products sold on Amazon's website collected by 2 million users. The rating includes 1–5. The larger the value, the more satisfied the user is with the product. We use its subset, *i.e.,* the rating of 886 products by 9697 users. The book dataset was collected by Cai Nicolas Ziegler on the book crossing community for 4 weeks. It contains 278858 users' scores on 271379 books. We used its subset, *i.e.,* evaluation of 150 books by 671 users.

In addition to the above four data sets, we also use five recommended algorithms to compare with the proposed algorithm. They are introduced as follows:

Usercf [43]: this method is a traditional user-based recommendation algorithm. Given a target user, it finds users similar to the target user from the known user set, and then recommends items for target user by using the hobbies of similar known users according to the measurement function.

Itemcf [43]: this method is similar to Usercf. The only difference is that it is based on the content information of the item and does not need to be based on the user's evaluation opinions on the item. *i.e.,* it needs to calculate the similarity between items, and recommend the items with high similarity to the historical items of the target user according to the measurement function.

LLAE [44]: it mainly aims at the cold start problem in the recommendation algorithm. Specifically, it first uses a low rank encoder to map the behavior space to the feature space, so that new users can connect to old users. Then, the symmetrical decoder reconstructs the user's behavior with the user's characteristics, so as to make recommendations for the target user.

JCA [45]: it is a hybrid recommendation algorithm. Specifically, it first learns the similarities between users and items at the same time. Then it uses a mini batch optimization algorithm to train the model. Finally, pairwise hinge based objective function is used to implicitly recommend items for target users.

XPL_CF [46]: this method is a feature-based collaborative filtering recommendation algorithm. It uses prior knowledge to encode each user's embedding into a sparse linear combination of item embedding. In addition, it can also mine the relationships between items.

### 4.2. Experimental setting

The experimental environment of this paper is windows 10, 64 bit operating system. All the codes are run on MATLAB 2016b soft-

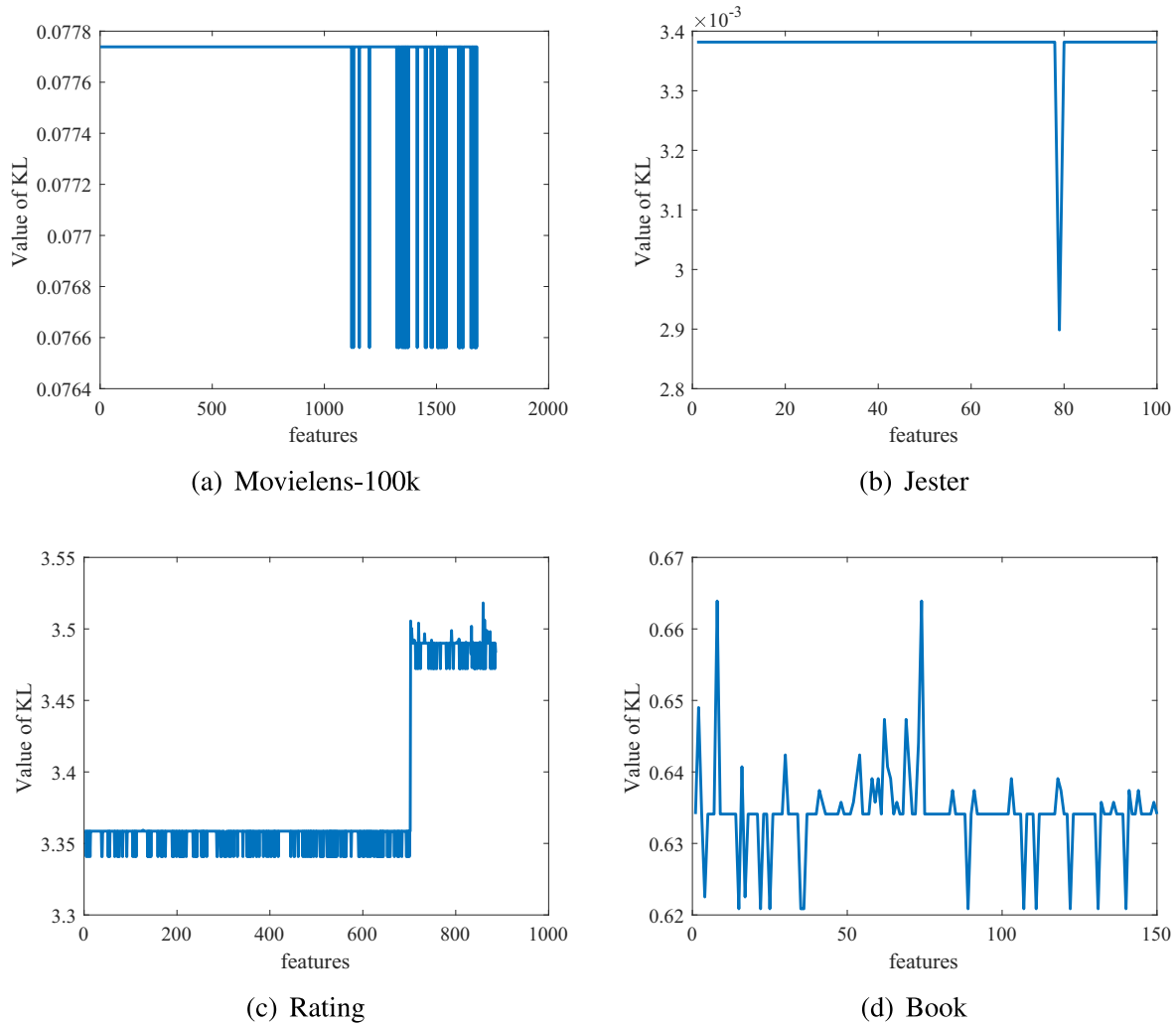

(a) Movielens-100k  (b) Jester

(c) Rating  (d) Book

**Fig. 14.** The KL value of each feature on all datasets.

214

ware. The datasets include Movielens-100 k, Jester, Ratings and Book. The comparison algorithms include Usercf, Itemcf, LLAE, JCA and XPL_ CF. Usercf is a collaborative filtering recommendation algorithm based on user similarity, Itemcf is a content-based recommendation algorithm, and JCA is a hybrid recommendation algorithm. LLAE is the recommended algorithm to solve the cold start problem. XPL_CF is a collaborative filtering recommendation algorithm based on data sparse representation. Neighbor relationship [42] is used to obtain the cross entropy, KL divergence and JS divergence of features, and then our method can work.

In our experiment, we use ten fold cross validation [47] to divide the training set and the test set (*i.e.,* divide the data set into 10 parts on average, of which 9 are the training set and 1 is the test set until all the data are tested). In the comparison experiment with the comparison algorithm, we use RMSE, MAE and running time as evaluation indexes. The value of RMSE is calculated by the following formula:

$$RMSE = \sqrt{\frac{1}{n}\sum_{u=1}^{n}(r_u - \hat{r}_u)^2} \quad (15)$$

where $n$ represents the number of scores, $r_u$ represents the user's real score, and $\hat{r}_u$ represents the predicted score given by the recommendation algorithm. MAE value is calculated by the following formula:

$$MAE = \frac{1}{n}\sum_{u=1}^{n}|r_u - \hat{r}_u| \quad (16)$$

Because the hyper-class provides a recommendation orientation for test users, it reduces the search range of similar users, and improves the speed and accuracy of recommendation. *i.e.,* our proposed hyper-class is to serve collaborative filtering recommendation. Therefore, we also did the parameter sensitivity experiment of our own algorithm (with the change of the nearest neighbor number $K$ and each fold of the ten fold cross validation of the test user, and the change of the values of RMSE and MAE).

### 4.3. Analysis of experimental results

#### 4.3.1. Comparative analysis

From Fig. 5, we can see that the hyper-class based algorithm CF_CE, CF_KL and CF_JS achieves better results than other algorithms. The reason is that CF_CE, CF_KL and CF_JS algorithms exclude most known users that are not related to the target user according to the hyper-class defined by us in advance. The Usercf algorithm needs to filter from all known users, and its possibility of finding known users that are not related to the target user is greatly improved. Similarly, Itemcf algorithm also looks for similarity from all items, which is easy to cause false recommendation to target users. LLAE and JCA improve the recommendation algo-rithm from the cold start and hybrid recommendation in the recommendation system respectively. XPL_CF improves the recommendation algorithm by learning the sparse representation of data. They have achieved certain results, but they still do not exclude some known users or items that are irrelevant or misleading to the prediction in advance.

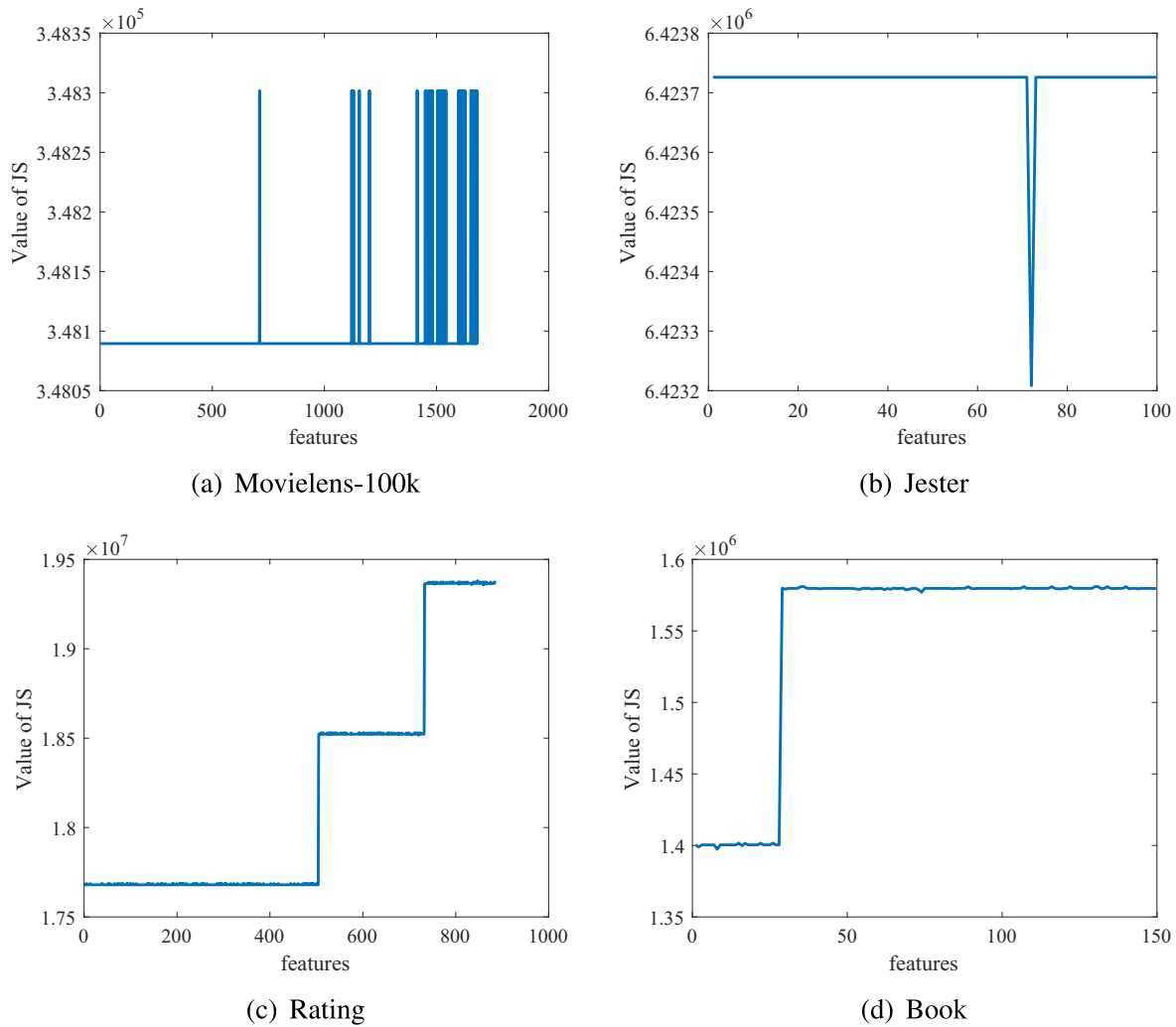

Fig. 15. The JS value of each feature on all datasets.

Fig. 6 shows the MAE results of all algorithms. Similarly, hyper-class based algorithm CF_CE, CF_KL and CF_JS achieves better results than other algorithms. The hyper-class we established is equivalent to a rough screening, which can eliminate the known users who are not related to the target users, so as to improve the accuracy of recommendation. On the Book dataset, itemcf algorithm achieves the best effect. The reason is that different datasets have different structural characteristics, and itemcf algorithm based on item similarity is more suitable for this data set. It should be noted that although the itemcf algorithm achieves the best Mae results on the Book dataset, the RMSE results on the Book dataset are not the best.

Table 2 and Table 3 show the average values of RMSE and MAE of all algorithms on four data sets respectively. From Table 2, compared with the worst Usercf algorithm, we can see the hyper-class based algorithm CF_CE, CF_KL and CF_JS have an average reduction of 2.3377, 2.2917 and 2.2853 in RMSE value, respectively. Compared with the best comparison algorithm XPL_CF, CF_CE, CF_KL and CF_JS decreased by an average of 0.2939, 0.2479 and 0.2415 in RMSE values, respectively. From Table 3, we can see the hyper-class based algorithm CF_CE, CF_KL and CF_JS compared with the worst Usercf algorithm, reduces the MAE value by 2.1097, 2.0465 and 2.0207 respectively. Compared with the best comparison algorithm JCA, CF_CE, CF_KL and CF_JS decreased by 0.6558, 0.5926 and 0.5668 in MAE value on average, respectively. It shows that the hyper-class we constructed is helpful to improve the recommendation algorithm, and it can improve the accuracy of the recommendation algorithm.

Table 4 shows the running costs of all algorithms. From Table 4, we can see that compared with the best comparison algorithm XPL_CF, our proposed CF_CE, CF_KL and CF_JS algorithm reduces the time by 61.49%, 62.21% and 63.14% respectively. The main reason for this phenomenon is that the hyper-class we built has given a fuzzy classification to the data. For a target user, it first finds the nearest class, and then recommends items for it according to the hobbies of the known users in the nearest class, which greatly reduces the search scope for the target user, thus speeding up the recommendation speed.

*4.3.2. Parameter sensitivity analysis*

CF_CE, CF_KL and CF_JS algorithm are improved collaborative filtering recommendation algorithms through the hyper-class established by using the defined cross entropy, KL divergence and JS divergence. Therefore, for a target user, we select *K* similar known users to recommend item for it. At the same time, ten fold cross validation method is used in this experiment. Based on this, we have done parameter sensitivity experiments for CF_CE, CF_KL and CF_JS three algorithms, as shown in Figs. 7–12. From Figs. 7–9, we can see that for CF_CE algorithm, in addition to Jester data set, the number of cross validation has a greater impact on it than k value. For CF_KL and CF_JS algorithm, whether the number of cross validation or *K* value, has a great impact on it, especially on book and rating data sets.

In Figs. 10–12, we can see the MAE values of three algorithms CF_CE, CF_KL and CF_JS varying with parameters on four datasets. Specifically, both the number of cross validation and the K value have an impact on them. Especially on the Movielens-100 k, Rating and Book datasets, some values are much higher than others. The reasons for this phenomenon are: 1. Different K values indicate that the number of similar known users of the target user is different, which will affect the recommendation of the target user. 2. Cross validation has randomness, and its data division is often random, which will affect the experimental results to a certain extent.

To sum up, algorithms CF_CE, CF_KL and CF_JS are sensitive to parameters. We need to select appropriate parameters to run them.

*4.3.3. Selected decision features*

Figs. 13–15 show the cross entropy, KL divergence and JS divergence values of each feature on four datasets. Specifically, on the Movielens-100 k dataset, CF_ CE selects the 50th feature as the decision feature, CF_ KL selects the 1122nd feature as the decision feature, CF_ JS selects the first feature as the decision feature. On the Jester dataset, CF_ CE selects the 79th feature as the decision feature, CF_ KL also selects the 79th feature as the decision feature, while CF_ JS selects the 72nd feature as the decision feature. This shows that Jester's 79th feature is most suitable for establishing hyper-class. On the Rating dataset, CF_ CE selects the 705th feature as the decision feature, CF_ KL selects the 175th feature as the decision feature, CF_ JS selects the 127th feature as the decision feature. On the book dataset, CF_ CE selects the 15th feature as the decision feature, CF_ KL also selects the 15th feature as the decision feature, while CF_ JS selects the 8th feature as the decision feature. It shows that in the Book dataset, the 15th feature is suitable to be used to build hyper-class.

In addition, we can also see that on some data sets, the cross entropy, KL divergence or JS divergence of multiple features have the smallest value. For example, the cross entropy and KL divergence on the Book dataset, the KL divergence on the Rating dataset and the Movielens-100 k dataset, and the JS value on the Movielens-100 k dataset. This shows that there are multiple features on these datasets that are suitable to be constructed as hyper-class. It should be noted that in this paper, we only select the feature corresponding to the first minimum cross entropy, KL divergence value or JS divergence value to construct the hyper-class.

## 5. Conclusion

This paper has put forward a hyper-class representation of data, named as HC representation. Hyper-class representation can help us utilize big data better and make data mining algorithms faster and more accurate. Specifically, the cross entropy, KL divergence and JS divergence of features in data are comprehensively considered in constructing the HC representation of data. HC representation is equivalent to the class annotation of data. It assists in understanding training data by clustering attribute values and ordering the correlation between attributes and the decision attribute in training dataset. A recommended algorithm is applied to verify its performance. *i.e.,* Given a target user, we first find the nearest class according to the hyper-class representation, and then recommend the item for test user according to the hobbies of the known users in the nearest class. It greatly reduces the search scope of the recommendation algorithm and finds similar potential users more accurately. In the experiment, compared with the state-of-the-art algorithms, the proposed algorithm achieves faster and better recommendation.

In the future work, we plan to study mainly two aspects as follows.

1. Exploring more suitable methods to establish hyper-class representation and applying it to other data mining algorithms, such as deep learning and SVM.
2. Addressing the unfairness, or limitations in attribute-level operations of data, and seeking new solutions in attribute–value-level of data.

**Declaration of Competing Interest**

The authors declare that they have no known competing financial interests or personal relationships that could have appeared to influence the work reported in this paper.

**Acknowledgement**

This work has been supported in part by the Natural Science Foundation of China under grant 61836016, and Research Fund of Guangxi Key Lab of Multi-source Information Mining and Security (MIMS20-04).

**References**

[1] L. Maier-Hein, M. Eisenmann, D. Sarikaya, K. März, T. Collins, A. Malpani, J. Fallert, H. Feussner, S. Giannarou, P. Mascagni, et al., Surgical data science–from concepts toward clinical translation, Medical Image Analysis 76 (2022) 102306.
[2] R. Rossi, K. Hirama, Characterizing big data management, arXiv preprint arXiv:2201.05929.
[3] J.P. Meier-Kolthoff, J.S. Carbasse, R.L. Peinado-Olarte, M. Göker, Tygs and lpsn: a database tandem for fast and reliable genome-based classification and nomenclature of prokaryotes, Nucleic Acids Research 50 (D1) (2022) D801–D807.
[4] P.K.D. Pramanik, S. Pal, M. Mukhopadhyay, Healthcare big data: A comprehensive overview, Research Anthology on Big Data Analytics, Architectures and Applications (2022) 119–147.
[5] M. Naeem, T. Jamal, J. Diaz-Martinez, S.A. Butt, N. Montesano, M.I. Tariq, E. De-la Hoz-Franco, E. De-La-Hoz-Valdiris, Trends and future perspective challenges in big data, in: Advances in Intelligent Data Analysis and Applications, Springer, 2022, pp. 309–325.
[6] Y. Zhao, S. Zhang, Generalized dimension-reduction framework for recent-biased time series analysis, IEEE Transactions on Knowledge and Data Engineering 18 (2) (2005) 231–244.
[7] S. Zhang, Challenges in knn classification, IEEE Transactions on Knowledge and Data Engineering (2021), https://doi.org/10.1109/TKDE.2021.3049250, 1–1.
[8] S. Zhang, Cost-sensitive knn classification, Neurocomputing 391 (2020) 234–242.
[9] S. Zhang, J. Li, Knn classification with one-step computation, IEEE Transactions on Knowledge and Data Engineering (2021), https://doi.org/10.1109/TKDE.2021.3119140, 1–1.
[10] S. Zhang, X. Li, M. Zong, X. Zhu, R. Wang, Efficient knn classification with different numbers of nearest neighbors, IEEE Transactions on Neural Networks and Learning Systems 29 (5) (2017) 1774–1785.
[11] S. Xie, T. Yang, X. Wang, Y. Lin, Hyper-class augmented and regularized deep learning for fine-grained image classification, in: Proceedings of the IEEE conference on computer vision and pattern recognition, 2015, pp. 2645–2654.
[12] A. Singleton, D. Arribas-Bel, Geographic data science, Geographical Analysis 53 (1) (2021) 61–75.
[13] H. Liu, X. Wu, S. Zhang, Neighbor selection for multilabel classification, Neurocomputing 182 (2016) 187–196.
[14] H. Jomard, O. Scotti, S. Auclair, P. Dominique, K. Manchuel, D. Sicilia, The sisfrance database of historical seismicity. state of the art and perspectives, Comptes Rendus. Géoscience 353 (S1) (2021) 1–24.
[15] H.E. Samra, A.S. Li, B. Soh, M.A. AlZain, Review of contemporary database design and implication for big data, International Journal of Smart Education and Urban Society (IJSEUS) 12 (4) (2021) 1–11.
[16] S. Zhang, Shell-neighbor method and its application in missing data imputation, Applied Intelligence 35 (1) (2011) 123–133.
[17] S. Zhang, D. Cheng, M. Zong, L. Gao, Self-representation nearest neighbor search for classification, Neurocomputing 195 (2016) 137–142.
[18] H.K. Sharma, T. Choudhury, R. Tomar, J. Patni, J.-S. Um, An algorithmic approach for performance tuning of a relational database system using dynamic sga parameters, Spatial Information Research (2021) 1–17.
[19] R. Mama, M. Machkour, K. Ahkouk, K. Majhadi, Towards a flexible relational database query system, in: Proceedings of the 4th International Conference on Networking, Information Systems & Security, 2021, pp. 1–5.
[20] L.P. Nguyen, et al., Exploring learned join algorithm selection in relational database management systems, Ph.D. thesis, Massachusetts Institute of Technology (2021).
[21] C. Lv, D. Pan, Y. Li, J. Li, Z. Wang, A novel chinese entity relationship extraction method based on the bidirectional maximum entropy markov model, Complexity (2021).
[22] G. Wu, J. Lin, C.T. Silva, Era: Entity relationship aware video summarization with wasserstein gan, arXiv preprint arXiv:2109.02625.
[23] S. Al-Fedaghi, Conceptual data modeling: Entity-relationship models as thinging machines, arXiv preprint arXiv:2109.14717.
[24] D. Thomas, B. Pagurek, R. Buhr, Validation algorithms for pointer values in dbtg databases, ACM Transactions on Database Systems (TODS) 2 (4) (1977) 352–369.
[25] D. Hawley, J. Knowles, E.E. Tozer, Database consistency and the codasyl dbtg proposals, The Computer Journal 18 (3) (1975) 206–212.
[26] Z. Deng, X. Zhu, D. Cheng, M. Zong, S. Zhang, Efficient knn classification algorithm for big data, Neurocomputing 195 (2016) 143–148.
[27] Y. Zhou, W. Zhang, J. Kang, X. Zhang, X. Wang, A problem-specific non-dominated sorting genetic algorithm for supervised feature selection, Information Sciences 547 (2021) 841–859.
[28] Y. Lu, X. Chen, Joint feature weighting and adaptive graph-based matrix regression for image supervised feature selection, Signal Processing: Image Communication 90 (2021) 116044.
[29] X. Chen, Y. Lu, J. Zhang, X. Zhu, Margin-based discriminant embedding guided sparse matrix regression for image supervised feature selection, Computer Vision and Image Understanding 212 (2021) 103273.
[30] C. Tang, X. Zheng, X. Liu, W. Zhang, J. Zhang, J. Xiong, L. Wang, Cross-view locality preserved diversity and consensus learning for multi-view unsupervised feature selection, IEEE Transactions on Knowledge and Data Engineering.
[31] R. Shang, L. Wang, F. Shang, L. Jiao, Y. Li, Dual space latent representation learning for unsupervised feature selection, Pattern Recognition 114 (2021) 107873.
[32] J. Miao, T. Yang, L. Sun, X. Fei, L. Niu, Y. Shi, Graph regularized locally linear embedding for unsupervised feature selection, Pattern Recognition 122 (2022) 108299.
[33] J. Li, S. Zhang, L. Zhang, C. Lei, J. Zhang, Unsupervised nonlinear feature selection algorithm via kernel function, Neural Computing and Applications 32 (11) (2020) 6443–6454.
[34] X. Li, Y. Zhang, R. Zhang, Semisupervised feature selection via generalized uncorrelated constraint and manifold embedding, IEEE Transactions on Neural Networks and Learning Systems.
[35] X. Chen, R. Chen, Q. Wu, F. Nie, M. Yang, R. Mao, Semisupervised feature selection via structured manifold learning, IEEE Transactions on Cybernetics.
[36] C. Wang, X. Chen, G. Yuan, F. Nie, M. Yang, Semisupervised feature selection with sparse discriminative least squares regression, IEEE Transactions on Cybernetics.
[37] N. Zeng, Z. Wang, W. Liu, H. Zhang, K. Hone, X. Liu, A dynamic neighborhood-based switching particle swarm optimization algorithm, IEEE Transactions on Cybernetics.
[38] X. Luo, Y. Yuan, S. Chen, N. Zeng, Z. Wang, Position-transitional particle swarm optimization-incorporated latent factor analysis, IEEE Transactions on Knowledge and Data Engineering.
[39] W. Yue, Z. Wang, J. Zhang, X. Liu, An overview of recommendation techniques and their applications in healthcare, IEEE/CAA Journal of Automatica Sinica 8 (4) (2021) 701–717.
[40] Y. Qian, J. Liang, W. Pedrycz, C. Dang, Positive approximation: an accelerator for attribute reduction in rough set theory, Artificial Intelligence 174 (9–10) (2010) 597–618.
[41] J. Liang, Z. Shi, D. Li, M.J. Wierman, Information entropy, rough entropy and knowledge granulation in incomplete information systems, International Journal of General Systems 35 (6) (2006) 641–654.
[42] Q. Hu, W. Pedrycz, D. Yu, J. Lang, Selecting discrete and continuous features based on neighborhood decision error minimization, IEEE Transactions on Systems, Man, and Cybernetics, Part B (Cybernetics) 40 (1) (2009) 137–150.
[43] J. Lu, D. Wu, M. Mao, W. Wang, G. Zhang, Recommender system application developments: a survey, Decision Support Systems 74 (2015) 12–32.
[44] J. Li, M. Jing, K. Lu, L. Zhu, Y. Yang, Z. Huang, From zero-shot learning to cold-start recommendation, in: Proceedings of the AAAI Conference on Artificial Intelligence, Vol. 33, 2019, pp. 4189–4196.
[45] Z. Zhu, J. Wang, J. Caverlee, Improving top-k recommendation via jointcollaborative autoencoders, in: The World Wide Web Conference, 2019, pp. 3483–3482.
[46] F.M. Almutairi, N.D. Sidiropoulos, B. Yang, Xpl-cf: Explainable embeddings for feature-based collaborative filtering, in: Proceedings of the 30th ACM International Conference on Information & Knowledge Management, 2021, pp. 2847–2851.
[47] G. Jiang, W. Wang, Markov cross-validation for time series model evaluations, Information Sciences 375 (2017) 219–233.

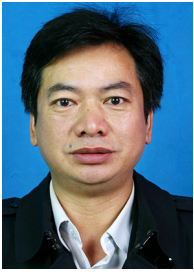

**Shichao Zhang** is a China National Distinguished Professor with the Central South University, China. He holds a PhD degree from the Deakin University, Australia. His research interests include data mining and big data. He has published 90 international journal papers and over 70 international conference papers. He is a CI for 18 competitive national grants. He is a senior member of the IEEE, a member of the ACM; and serves/served as an associate editor for four journals.

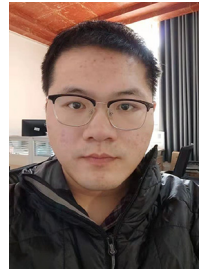

**Wenzhen Zhang** is currently working toward the PhD degree at Guangxi Normal University, China. His research interests include machine learning, data mining and deep learning.

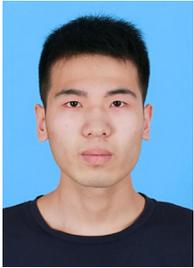

**Jiaye Li** is currently working toward the PhD degree at Central South University, China. His research interests include machine learning, data mining and deep learning

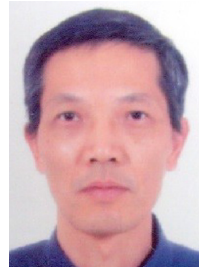

**Yongsong Qin** is professor at Guangxi Normal University, China. He holds a PhD degree from the University of Science and Technology of China. His research interests include statistical inference and data mining. He has published 60 international journal papers.